\documentclass[11pt,a4paper]{article}
\pdfoutput=1 

\usepackage[table]{xcolor}
\usepackage{colortbl}
\RequirePackage{ifpdf} 
\usepackage{amsmath} 
\usepackage{mathtools}
\usepackage{cases}

\usepackage{jheppub}
\usepackage{pstricks}
\usepackage[final]{pdfpages} 
\usepackage{ifpdf} 
\usepackage{slashed}

\usepackage{color}
\usepackage{xcolor}
\definecolor{urlblue}{rgb}{0.2,0.4,0.7}
\definecolor{citegreen}{rgb}{0,0.6,0.2}
\definecolor{linkred}{rgb}{0.9,0.2,0.1}
\usepackage{hyperref}
\hypersetup{
colorlinks=true, citecolor=citegreen, linkcolor=blue, urlcolor=urlblue}

\usepackage{graphics}
\usepackage{etoolbox} 
\usepackage{fixmath}
\usepackage{psfrag}

\usepackage{notoccite} 

\usepackage{amsfonts}
\usepackage{autobreak}
\usepackage{marginnote}

\usepackage{tikz}
\usetikzlibrary{positioning,arrows}
\usetikzlibrary{decorations.pathmorphing}
\usetikzlibrary{decorations.markings}
\usetikzlibrary{shapes.geometric}
\tikzset{
    vector/.style={decorate, decoration={snake}, draw},
    provector/.style={decorate, decoration={snake,amplitude=2.5pt}, draw},
    antivector/.style={decorate, decoration={snake,amplitude=-2.5pt}, draw},
    fermion/.style={draw=black, postaction={decorate},decoration={markings,mark=at position .55 with {\arrow[draw=black]{>}}}},
    fermionbar/.style={draw=black, postaction={decorate},
                       decoration={markings,mark=at position .55 with {\arrow[draw=black]{<}}}},
    fermionnoarrow/.style={draw=black},
    gluon/.style={decorate, draw=black,decoration={coil,amplitude=4pt, segment length=5pt}},
    scalar/.style={dashed,draw=black, postaction={decorate},decoration={markings,mark=at position .55 with {\arrow[draw=black]{>}}}},
    scalarbar/.style={dashed,draw=black, postaction={decorate},decoration={markings,mark=at position .55 with {\arrow[draw=black]{<}}}},
    scalarnoarrow/.style={dashed,draw=black},
    electron/.style={draw=black, postaction={decorate},decoration={markings,mark=at position .55 with {\arrow[draw=black]{>}}}},
    bigvector/.style={decorate, decoration={snake,amplitude=4pt}, draw},
}

\title{The Curious Case of Leading Transcendentality: three-point Form Factors}

\author{Taushif Ahmed$^{a}$, Pulak Banerjee$^{b}$, Amlan Chakraborty$^{c}$, Prasanna K. Dhani$^{d}$ and V. Ravindran$^{c}$}
\emailAdd{taushif@mpp.mpg.de,
pulak.banerjee@psi.ch,
amlanchak@imsc.res.in, 
dhani@fi.infn.it,
ravindra@imsc.res.in}

\affiliation{$^a$Max-Planck-Institut f\"ur Physik, Werner-Heisenberg-Institut, 80805 M\"unchen, Germany \\
$^b$Paul Scherrer Institut, Forschungsstrasse 111, CH-5232 Villigen PSI, Switzerland\\
$^c$The Institute of Mathematical Sciences, HBNI, Taramani, Chennai 600113, India\\
$^d$INFN, Sezione di Firenze, I-50019 Sesto Fiorentino, Florence, Italy}

\preprint{IMSc/2019/0506, MPP-2019-100, PSI-PR-19-10}

\abstract{Form factors are important ingredients to investigate the principle of maximal transcendentality (PMT) and to extract anomalous dimensions of local gauge invariant operators. In this article, we compute several two- and three-point FFs to three- and two-loops, respectively, for three different choices of local gauge invariant operators in ${\cal N}=4$ super Yang-Mills (SYM) theory. The operators ${\cal O}^1$ and ${\cal O}^2$ are flavour and helicity blind configurations of supersymmetric descendant of the half-BPS and Konishi primary, respectively, and ${\cal O}^3$ is the energy-momentum tensor. The operators ${\cal O}^1$ and ${\cal O}^2$ are composed of non-protected  dimension-three (classical) fermionic and scalar components belonging to SU($2|3$) closed sub-sector of ${\cal N}=4$ SYM. We analyse the mixing among the non-protected fermionic and scalar components of these operators up to three-loops in perturbation theory and consequently, compute the quantum corrections to the corresponding dilatation operators. The highest transcendental (HT) weight terms of the FFs of ${\cal O}^1$ are found to be independent of the external on-shell states and, moreover, those are equal to that of half-BPS, however, this does not hold true for the FFs of ${\cal O}^2$. FFs of the ${\cal O}^3$ exhibit identical behaviours to that of half-BPS, in concordance with the classical expectations. However, the three-point FFs of ${\cal O}^3$  violate the PMT while comparing with the corresponding quantity in the standard model, observed for the first time at the level of FF.}

\begin{document}
\allowdisplaybreaks[4]
\unitlength1cm
\maketitle
\flushbottom


\def\D{{\cal D}}
\def\DD{\overline{\cal D}}
\def\g{\overline{\cal C}}
\def\gm{\gamma}
\def\M{{\cal M}}
\def\ep{\epsilon}
\def\epm1{\frac{1}{\epsilon}}
\def\epm2{\frac{1}{\epsilon^{2}}}
\def\epm3{\frac{1}{\epsilon^{3}}}
\def\epm4{\frac{1}{\epsilon^{4}}}
\def\unM{\hat{\cal M}}
\def\ashat{\hat{a}_{s}}
\def\asmur{a_{s}^{2}(\mu_{R}^{2})}
\def\sigbar{{{\overline {\sigma}}}\left(a_{s}(\mu_{R}^{2}), L\left(\mu_{R}^{2}, m_{H}^{2}\right)\right)}
\def\sigbarn{{{{\overline \sigma}}_{n}\left(a_{s}(\mu_{R}^{2}) L\left(\mu_{R}^{2}, m_{H}^{2}\right)\right)}}
\def\unas{ \left( \frac{\hat{a}_s}{\mu_0^{\epsilon}} S_{\epsilon} \right) }
\def\rnM{{\cal M}}
\def\bt{\beta}
\def\cD{{\cal D}}
\def\cC{{\cal C}}
\def\ca{\text{\tiny C}_\text{\tiny A}}
\def\cf{\text{\tiny C}_\text{\tiny F}}
\def\ct{{\red []}}
\def\sv{\text{SV}}
\def\murOmu{\left( \frac{\mu_{R}^{2}}{\mu^{2}} \right)}
\def\bb{b{\bar{b}}}
\def\bt0{\beta_{0}}
\def\bt1{\beta_{1}}
\def\bt2{\beta_{2}}
\def\bt3{\beta_{3}}
\def\gm0{\gamma_{0}}
\def\gm1{\gamma_{1}}
\def\gm2{\gamma_{2}}
\def\gm3{\gamma_{3}}
\def\nn{\nonumber}
\def\l{\left}
\def\r{\right}
\def\T{{\cal Z}}    
\def\U{{\cal Y}}

\def\nn{\nonumber\\}
\def\ep{\epsilon}
\def\T{\mathcal{T}}
\def\V{\mathcal{V}}

\def\qgraf{{\fontfamily{qcr}\selectfont
QGRAF}}
\def\python{{\fontfamily{qcr}\selectfont
PYTHON}}
\def\form{{\fontfamily{qcr}\selectfont
FORM}}
\def\reduze{{\fontfamily{qcr}\selectfont
REDUZE2}}
\def\kira{{\fontfamily{qcr}\selectfont
Kira}}
\def\litered{{\fontfamily{qcr}\selectfont
LiteRed}}
\def\fire{{\fontfamily{qcr}\selectfont
FIRE5}}
\def\air{{\fontfamily{qcr}\selectfont
AIR}}
\def\mint{{\fontfamily{qcr}\selectfont
Mint}}
\def\hepforge{{\fontfamily{qcr}\selectfont
HepForge}}
\def\arXiv{{\fontfamily{qcr}\selectfont
arXiv}}
\def\Python{{\fontfamily{qcr}\selectfont
Python}}
\def\ginac{{\fontfamily{qcr}\selectfont
GiNaC}}
\def\polylogtools{{\fontfamily{qcr}\selectfont
PolyLogTools}}
\def\anci{{\fontfamily{qcr}\selectfont
Finite\_ppbk.m}}

\newcommand{\dis}{}
\newcommand{\overbar}[1]{mkern-1.5mu\overline{\mkern-1.5mu#1\mkern-1.5mu}\mkern
1.5mu}


\section{Introduction}
\label{sec:intro}
On-shell scattering amplitudes and off-shell correlation functions are quantities of fundamental importance in any gauge theory. A generic quantum field theory is completely specified by the knowledge of these quantities. Form factors (FFs), the overlap of an $n$-particle on-shell state with a state created by the action of a local gauge invariant operator on the vacuum, are a fascinating bridge between aforementioned two quantities which have been studied extensively in quantum chromodynamics (QCD) and ${\cal N}=4$ super Yang-Mills (SYM) theory:
\begin{align}
    {\cal F}_{\cal O}(1,\ldots,n;q) \equiv \langle 1,\ldots,n|{\cal O}(0)|0\rangle\,.
\end{align}
The numbers $1,\ldots,n$ denote the on-shell particles and $q^2\neq 0$ is the off-shell momentum associated with the operator. The central objects which are considered in this article are two- and three-point FFs. Very recently, the first calculations of generalisations of such FFs to the case of two operator insertions with one non-protected operator was performed~\cite{Ahmed:2019upm} by some of us.

In this work, we begin our discussion by considering two local gauge invariant operators:
\begin{alignat}{2}
\label{eq:op-def}
  &  {\cal O}^1&&={\bar \lambda}^a_m  {\lambda}^a_m + 4\,g\,f^{abc} \varepsilon_{ijk} \phi^a_i \phi^b_j \phi^c_k \,,\nonumber\\
  &  {\cal O}^2&&= f^{abc} \varepsilon_{ijk} \phi^a_i \phi^b_j \phi^c_k    -\,\frac{g}{4}\ {\bar \lambda}^a_m  {\lambda}^a_m \,.
\end{alignat}
The Majorana and scalar fields are denoted through $\lambda$ and $\phi$, respectively. All the fields in ${\cal N}=4$ SYM transform under the adjoint representation of SU(N) gauge group which is captured through the indices $a,b,c$ with $\delta_{aa}=N^2-1$. The Yang-Mills coupling strength is denoted by $g$. Four generations of Majorana fermions are represented through $m,n \in [1,4]$. The indices $i,j \in [1,n_g]$ denote the generation number of scalar and pseudo-scalar fields with $n_g=3$ in 4-dimensions. The fully anti-symmetric structure constants of SU(N) group is defined through $(T^a)_{bc}=-if^{abc}$. 
The aforementioned operators in \eqref{eq:op-def} are constructed out of non-protected, dimension-three (classical) fermionic and scalar components which belong to SU$(2|3)$ closed subsector of ${\cal N}=4$ SYM. The FFs in this sector were first studied in ref.~\cite{Brandhuber:2016fni}. The helicity and flavour dependent counterparts of the operator ${\cal O}^1$ was obtained~\cite{Intriligator:1999ff,Bianchi:2001cm,Beisert:2003ys} as conformal descendant of the half-BPS and the ${\cal O}^2$ was obtained as supersymmetric descendant of Konishi primary~\cite{Eden:2005ve,Brandhuber:2016fni}. 

Understanding the mathematical structures of scattering amplitudes and FFs in ${\cal N}=4$ SYM has been an active area of investigation for past several decades, not only due to its underlying ultraviolet conformal symmetry but also its ability to help in solving more complicated theory like QCD. One of the most interesting facts is the presence of uniform transcendental\footnote{The weight of transcendentality, $\tau$, of a function, $f$, is defined as the number of iterated integrals required to define the function $f$, e.g. $\tau(\log)=1\,, \tau({\rm Li}_n)=n\,, \tau(\zeta_n)=n$ and also we define $\tau(f_1f_2)=\tau(f_1)+\tau(f_2)$. Algebraic factors are assigned weight zero and dimensional regularisation parameter $\epsilon$ to -1.} (UT) terms in certain class of scattering amplitudes and FFs in ${\cal N}=4$ SYM. This is indeed an observational~\cite{Kotikov:2002ab,Kotikov:2004er,Bern:2006ew,Drummond:2007cf,Naculich:2008ys,Bork:2010wf,Gehrmann:2011xn,Brandhuber:2012vm,Eden:2012rr,Drummond:2013nda,Basso:2015eqa,Ahmed:2016vgl,Goncalves:2016vir,Banerjee:2016kri,Banerjee:2018yrn}, albeit unproven, fact. The two-point or Sudakov FFs of the half-BPS operator belonging to the stress-energy supermultiplet is found~\cite{vanNeerven:1985ja,Brandhuber:2010ad,Gehrmann:2011xn} to contain terms with UT, more specifically, they are composed of only highest transcendental (HT) terms with weight 2L at loop order L. It is also shown in ref.~\cite{Gehrmann:2011xn} that it is possible to choose a representation in which each loop integral has uniform transcendentality. This has profound implications in choosing the basis of master integrals while computing Feynman loop integrals through method of differential equation. Therefore, it is very natural to examine how far this behaviour holds true. In light of this, in ref.~\cite{Brandhuber:2012vm}, the three-point FFs of the half-BPS operator are investigated and those are found to respect this wonderful property. On the contrary, this behaviour fails for the Konishi~\cite{KONISHI1984439}, the primary operator for the Konishi supermultiplet, for two-~\cite{Nandan:2014oga} as well as three-point~\cite{Banerjee:2016kri} FFs.  All the aforementioned results are in accordance with the general observation that the FFs of supersymmetry (SUSY) protected operators exhibit UT behaviour. However, there is no clear indication whether the reverse is also true. Through our computations, we find that the three-point one-loop FF of operator ${\cal O}^1$ with $\phi\phi\phi$ on-shell states does not exhibit UT behaviour. However, all the remaining two- and three-point FFs of ${\cal O}^1$ and ${\cal O}^2$ do not exhibit UT behaviours.

It is conjectured in ref.~\cite{Loebbert:2015ova} that at two-loop level the HT weight parts of every two-point minimal FF (number of fields present in the operator is same as number of external on-shell states) are identical and those are same as that of half-BPS operator belonging to the stress-energy supermultiplet~\cite{vanNeerven:1985ja}. In ref.~\cite{Ahmed:2016vgl}, this conjecture is verified to three-loops level for the FF of unprotected Konishi operator. In ref.~\cite{Brandhuber:2014ica}, the minimal FFs of long BPS operators (more than two fields) are computed to two-loops, and the corresponding HT piece is found to be a universal in all FFs with long, unprotected operators~\cite{Loebbert:2015ova,Brandhuber:2016fni,Brandhuber:2017bkg,Jin:2018fak,Jin:2019ile,Brandhuber:2018kqb}. The operators in \eqref{eq:op-def} are combination of long and short unprotected parts. Moreover, the corresponding FFs can neither be regarded as minimal nor non-minimal case. This kind of scenario is largely unexplored. Through our computations of three-point FFs, we show that the HT piece of ${\cal O}^1$ is independent of the external states and it is equal to that of half-BPS~\cite{Brandhuber:2012vm,Banerjee:2016kri}. However, for the operator ${\cal O}^2$, this property fails for all the FFs.

It is also conjectured in~\cite{Loebbert:2015ova}, that the HT terms of two-loop remainder function of the three-point FF of every length-two operator should agree with the corresponding half-BPS remainder found in ref.~\cite{Brandhuber:2012vm}. The latter conjecture is falsified in ref.~\cite{Banerjee:2016kri} where, for the first time, it is shown that for three-point FF of the Konishi operator (length-two), the HT part depends on the nature of external on-shell states; it fails to match with that of half-BPS if the external on-shell states are $\phi\lambda\lambda$. The operators in \eqref{eq:op-def} are combinations of length-two and three operators. As mentioned in the previous paragraph, though the HT parts of the ${\cal O}^1$ three-point FFs are identical to that of half-BPS~\cite{Brandhuber:2012vm,Banerjee:2016kri}, it fails for the FFs of ${\cal O}^2$ which can be thought of as a general feature of Konishi primary and its descendants.

In past few decades, people have been investigating the connection among quantities computed in different gauge theories. In particular, the connection between on-shell amplitudes or FFs of ${\cal N}=4$ SYM and that of QCD are of fundamental importance. Besides theoretical understanding, this is motivated from the fact that calculating any quantity in QCD is much more complex and in absence of our ability to calculate a quantity in QCD, whether it is possible to get the result, at least partially, from some other simpler theory. In refs.~\cite{Kotikov:2001sc,Kotikov:2004er,Kotikov:2006ts}, the connection between anomalous dimensions of leading twist-two operators of these theories is found and it is shown that the results in ${\cal N}=4$ SYM is related to the HT part of the QCD results and consequently, the principle of maximal transcendentality (PMT) is conjectured. Same conclusion is obtained by some of us in~\cite{Banerjee:2018yrn} through a different procedure based on momentum fraction space. At this level, this connection involves only pure number. This property is later examined in the context of two-point FFs in~\cite{Gehrmann:2011xn} to three-loops level and surprisingly, found to hold true: the HT pieces of quark (vector interaction) and gluon (scalar interaction) FFs in QCD~\cite{Gehrmann:2011aa} are identical to scalar FFs of half-BPS operator in ${\cal N}=4$ SYM upon changing the representation of fermions in QCD from fundamental to adjoint. The same behaviour is also found for the quark and gluon two-point FFs~\cite{Ahmed:2015qia,Ahmed:2016qjf} associated to tensorial interaction through energy-momentum tensor. In refs.~\cite{Brandhuber:2012vm,Banerjee:2017faz,Brandhuber:2017bkg,Jin:2018fak,Jin:2019ile}, the same behaviour is found to replicate for three-point scalar and pseudo-scalar FFs. This is the first scenario where non-trivial kinematics is involved and the validity of this principle implies this not only holds true for pure numbers but also for functions containing non-trivial kinematic dependence. Even after including the dimension seven operators in the effective theory of Higgs boson in the Standard model, PMT is also found to hold true~\cite{Jin:2018fak,Brandhuber:2018xzk,Jin:2019opr} for three-point FFs. Using this principle, the four-loop collinear anomalous dimension in planar ${\cal N}=4$ SYM is determined in ref.~\cite{Dixon:2017nat}. The complete domain of validity of this principle is still not fully clear and it is under active investigation. For on-shell scattering amplitudes, it breaks down even at one loop~\cite{Bern:1993mq} for cases with four or five external gluons. Considering three different operators, in this article, we check the validity of this conjecture for various two- and three-point FFs.


The aforesaid operators in \eqref{eq:op-def} are composed of two parts, one containing two Majorana fields, and the other with three scalar fields. Under quantum corrections, these components mix with each other and develop ultraviolet (UV) poles which manifest as non-zero anomalous dimensions. In order to investigate the mixing and determine the underlying anomalous dimensions, we compute several three-point two-loop and two-point three-loop form factors with the operators ${\cal O}^1$ and ${\cal O}^2$ insertions. By exploiting the universality of infrared divergences, we calculate the anomalous dimension matrix for the operator ${\cal O}^1$ fully up to two-loops and partially at three-loop, and for ${\cal O}^2$ partially to two-loops. Our findings of ${\cal O}^1$ are consistent with the 
partially two-loop results available in ref.~\cite{Brandhuber:2016fni}. The anomalous dimensions of ${\cal O}^2$ are identical to the Konishi supermultiplet, in accordance with the expectations~\cite{Eden:2005ve,Brandhuber:2016fni}. The anomalous dimensions are the corrections to the corresponding dilatation operators which can provide us non-trivial quantum field theoretic test for integrability.



One of the main motivations for constructing helicity, flavour blind operators in \eqref{eq:op-def}, in contrast to the ones considered in refs.~\cite{Intriligator:1999ff,Bianchi:2001cm,Beisert:2003ys,Eden:2005ve,Eden:2009hz,Brandhuber:2016fni} is to compare its FFs with known QCD results available for $H \rightarrow b\bar{b}g$ and $b\bar{b} \rightarrow H$ amplitudes directly, where $H,g,b$ stand for the Higgs boson, gluon and bottom quark, respectively.

Energy-momentum tensor (EM), which is also considered in this article, has been studied extensively in the context of three-point FFs in QCD~\cite{Ahmed:2014gla,Ahmed:2016yox}, and is found to behave like the half-BPS at one- as well as two-loop order and it is independent of the associated external on-shell states. This is in accordance with our classical expectation: since the stress-tensor is protected and lies in the same multiplet as the half-BPS, an exact SUSY Ward identity would relate these two FFs. Quite surprisingly, the process with three external partons violates the PMT even at one-loop while comparing the corresponding quantities in QCD! This is observed for the first time at the level of three-point FF.

After an initial introduction to the basics of ${\cal N}=4$ SYM, we present the explicit form of the EM tensor in section~\ref{sec:th}. All the processes that are analysed up to two- and three-loops level are also tabulated in the same section. In the following section~\ref{sec:FF}, we define the FFs and relate those with underlying scattering matrix elements. The methodology of the computation is briefly discussed in section~\ref{sec:calc}. In this article, we use the Feynman diagrammatic approach, in contrast to popular on-shell unitarity methods in the context of ${\cal N}=4$ SYM, along with the integration-by-parts~\cite{Tkachov:1981wb,Chetyrkin:1981qh} identities. Though ${\cal N}=4$ SYM is ultraviolet finite, that manifests through the vanishing $\beta$-function, nevertheless we need to regularise the theory not only because of the presence of soft and collinear divergences, but also due to appearance of non-zero anomalous dimensions for unprotected operators. In order to see these divergences explicitly and make sense out of the quantities we are interested in, we regularise the theory using modified dimensional reduction (${\overline{\rm DR}}$)~\cite{Siegel:1979wq,Capper:1979ns} which is discussed in section~\ref{sec:reg}. In the presence of unprotected operators, the method to perform UV renormalisation is discussed in section~\ref{ss:uv}. The universal behaviour of infrared (soft and collinear) divergences is described in section~\ref{ss:ir}. In section~\ref{ss:mixing}, the mixing among the constituent operators due to quantum corrections is demonstrated.  In addition to calculating the finite FFs following Catani's subtraction scheme, we also compute finite remainders which is introduced in section~\ref{sec:FinRem}. The behaviour of the FFs in the light of UT and PMT is analysed in details in section~\ref{sec:lt} which encapsulates many interesting findings. We also perform several consistency checks, discussed in section~\ref{sec:checks}, analytical as well as numerical to increase the reliability of our results. Concluding remarks are made in the last section~\ref{sec:concl}. In the appendix~\ref{app:results}, we give explicit analytic results of all the FFs.

\section{Theoretical framework}
\label{sec:th}

The dynamics of ${\cal N}=4$ SYM is encapsulated through the Lagrangian density~\cite{Brink:1976bc,Gliozzi:1976qd,
Jones:1977zr,Poggio:1977ma}
\begin{align}
    \mathcal{L}_{{\cal N} = 4} = &-\frac{1}{4}G_{\mu\nu}^a G^{\mu\nu a} - \frac{1}{2\xi}(\partial_{\mu}A^{a\mu})^2 + \partial_{\mu}\bar{\eta}^a 
D^{\mu}\eta_a + \frac{i}{2}\bar{\lambda}^a_m\gamma^{\mu}D_{\mu}\lambda^a_m + \frac{1}{2}(D_{\mu}\phi^a_i)^2 
\nonumber\\
&+ \frac{1}{2}(D_{\mu}\chi^a_i)^2 - \frac{g}{2}f^{abc}\bar{\lambda}^a_m[\alpha^i_{m,n}\phi^b_i +
 \gamma_5\beta^i_{m,n}\chi^b_i]\lambda^c_n - \frac{g^2}{4}\Big[(f^{abc}\phi^b_i\phi^c_j)^2 
\nonumber\\
&+ (f^{abc}\chi^b_i\chi^c_j)^2 + 2 (f^{abc}\phi^b_i\chi^c_j)^2\Big]\,.
\end{align}
The gauge, ghost, Majorana, scalar and pseudo-scalar fields are denoted through $A,\eta,\lambda,\phi$ and $\chi$, respectively. $G$ represents the field strength tensor that is related to the gauge field through covariant derivate $D$. All the fields in ${\cal N}=4$ SYM transform under the adjoint representation of SU(N) gauge group which is captured through the indices $a,b,c$ with $\delta_{aa}=N^2-1$. The Yang-Mills coupling strength is denoted by $g$ and $\xi$ is the gauge fixing parameter. Four generations of Majorana fermions are represented through $m,n \in [1,4]$. The indices $i,j \in [1,n_g]$ denote the generation numbers of scalar and pseudo-scalar fields with $n_g=3$ in 4-dimensions. The fully anti-symmetric structure constants of SU(N) group is defined through $(T^a)_{bc}=-if^{abc}$.
$\alpha$ and $\beta$ are anti-symmetric matrices satisfying the SUSY algebra:
\begin{align}
    [\alpha^i,\alpha^j]_+=[\beta^i,\beta^j]_+=-2\delta^{ij}, \quad [\alpha^i,\beta^j]_-=0
\end{align}
\begin{align}
    &{\rm tr}(\alpha^i)={\rm tr}(\beta^j)={\rm tr}(\alpha^i\beta^j)=0,\quad {\rm tr}(\alpha^i\alpha^j)={\rm tr}(\beta^i\beta^j)=-4\delta^{ij}\,,\nonumber\\
    &\qquad\qquad\quad {\rm tr}(\alpha^i\alpha^j\alpha^k)={\rm tr}(\beta^i\beta^j\beta^k)=-4\varepsilon_{ijk}\,.
\end{align}
In this article, in addition to considering the operators ${\cal O}^1$ and ${\cal O}^2$ defined in \eqref{eq:op-def}, we also consider the energy-momentum tensor which reads
\begin{alignat}{2}
\label{eq:op-def-EM}
    &{\cal O}^3_{\mu\nu}&&=\, G^{a}_{\mu \lambda} G^{\lambda}_{a \nu} + \frac{1}{4} 
\eta_{\mu\nu} G^{a}_{\rho \lambda} G^{\rho \lambda}_{a} - \frac{1}{\xi}\, \partial_{\lambda} A^{\lambda}
\left[ \partial_\mu A_\nu + \partial_\nu A_\mu  \right] + \frac{1}{2\xi}\, \eta_{\mu\nu} {(\partial_\rho {A^\rho}_a)}^2 \nn &{\color{white}xxx} &&+ (\partial_\mu \bar{\eta}^{a}) (D_{\nu} \eta_{a}) + (\partial_\nu \bar{\eta}^{a}) (D_{\mu} \eta_{a})
 - \eta_{\mu \nu} (\partial_\rho \bar{\eta}^{a}) (D^{\rho} \eta_{a})
 + \frac{i}{4} \Big[\bar{\lambda}^{a}_{m} \gamma_{\mu} D_{\nu} \lambda^{a}_{m} 
 \nn &{\color{white}xxx} && 
 + \bar{\lambda}^{a}_{m} \gamma_{\nu} 
 D_{\mu} \lambda^{a}_{m} - \frac{1}{2} \partial_{\mu} (
 \bar{\lambda}^{a}_{m} \gamma_{\nu} \lambda^{a}_{m})   
  -\frac{1}{2}  \partial_{\nu}   (\bar{\lambda}^{a}_{m} \gamma_{\mu} \lambda^{a}_{m} )\Big] 
  - \frac{i}{2} \Big[ \eta_{\mu \nu} \bar{\lambda}^{a}_{m} \gamma^{\rho} D_{\rho} \lambda^{a}_{m} 
  \nn &{\color{white}xxx} &&
  - \frac{1}{2} \eta_{\mu \nu} \partial_{\rho} \left( \bar{\lambda}^{a}_{m} \gamma^{\rho} \lambda^{a}_{m} \right) \Big]
  + (D_\mu \phi_{i}^{a}) (D_\nu \phi_{i}^{a}) - \frac{1}{2} \eta_{\mu \nu} (D_\rho \phi_{i}^{a})^{2}
  + (D_\mu \chi_{i}^{a}) (D_\nu \chi_{i}^{a})
 \nn &{\color{white}xxx} &&
 - \frac{1}{2} \eta_{\mu \nu} (D_\rho \chi_{i}^{a})^{2}  
+ \frac{g}{2} \eta_{\mu \nu} f^{abc} \,
 \bar{\lambda}_{m}^{a} \left[ \alpha^{i}_{m,n} \phi_{i}^{b} + \gamma_5 \beta^{i}_{m,n} \chi_{i}^{b}\right] \lambda_{n}^{c}
+\frac{g^2}{4} \eta_{\mu \nu} \Big[ (f^{abc} \phi^{b}_{i} \phi^{c}_{j})^2 
\nn &{\color{white}xxx} &&
+  (f^{abc} \chi^{b}_{i} \chi^{c}_{j})^2  
+ 2 (f^{abc} \phi^{b}_{i} \chi^{c}_{j} )^2\Big]\,.
\end{alignat}
This operator was considered in ref.~\cite{Banerjee:2018yrn} in computing all the splitting functions in $\mathcal{N}=4$ SYM. 
%
%
We are interested in two- and three-point form factors to three- and two-loops, respectively, consisting of a colourless off-shell state ($J^i$), described by the insertion of a local gauge invariant operator ${\cal O}^i$
\begin{align}
\label{eq:process}
    J^i(q) &\rightarrow 
    \begin{cases}
    &{\cal C}_1(p_1)+{\cal C}_2(p_2)\\
    &{\cal C}_1(p_1)+{\cal C}_2(p_2)+{\cal C}_3(p_3)\,,
    \end{cases}
\end{align}
where $p_i$ and $q$ are the corresponding 4-momentum with $p_i^2= 0, q^2\neq 0$. The outgoing on-shell particles are denoted by ${\cal C}_i=\{\phi, g, \lambda\}$. In this article, we consider several processes which are
\begin{align}
    J^1(q) &\rightarrow 
    \begin{cases}
    \label{eq:procO1-2pt}
    &\lambda(p_1)+\bar\lambda(p_2)\\
    \end{cases}\\
    J^1(q) &\rightarrow 
    \begin{cases}
    \label{eq:procS2}
    &g(p_1)+\lambda(p_2)+\bar\lambda(p_3)\\
    &\phi(p_1)+\lambda(p_2)+\bar\lambda(p_3) \\ 
    &\phi(p_1)+\phi(p_2)+\phi(p_3)
    \end{cases}\\
     J^2(q) &\rightarrow 
    \begin{cases}
    \label{eq:procS}
    &g(p_1)+\lambda(p_2)+\bar\lambda(p_3)\\
    &\phi(p_1)+\lambda(p_2)+\bar\lambda(p_3) \\ 
    &\phi(p_1)+\phi(p_2)+\phi(p_3)
    \end{cases}\\
    J^3(q) &\rightarrow 
    \begin{cases}
    \label{eq:procT}
    &g(p_1)+g(p_2)+g(p_3)\\
    &g(p_1)+\phi(p_2)+\phi(p_3)\\
    &g(p_1)+\lambda(p_2)+\bar\lambda(p_3)\\
    &\phi(p_1)+\lambda(p_2)+\bar\lambda(p_3)\,.
    \end{cases}
\end{align}
%
The underlying Lagrangian density encapsulating the interaction of the off-shell states described by gauge invariant local operators, defined in \eqref{eq:op-def}, to the fundamental fields of ${\cal N}=4$ SYM is given by
\begin{equation}
{\cal L}_{\rm int}^i = J^i {\cal O}^i.
\end{equation}
So, the full underlying Lagrangian for the scattering processes under consideration reads
\begin{align}
    {\cal L}^i={\cal L}_{{\cal N}=4}+{\cal L}^i_{\rm int}\,.
\end{align}
We define the kinematic invariants of the processes through
\begin{align}
\label{eq:mandel}
    s\equiv (p_1+p_2)^2\,,~~ t\equiv (p_2+p_3)^2\, ~~\text{and}~~ u\equiv (p_3+p_1)^2,
\end{align}
satisfying $s+t+u=q^2$. We also introduce dimensionless invariants out of these as
\begin{align}
\label{eq:dimless-var}
        x\equiv \frac{s}{q^2}\,,~~ y\equiv \frac{u}{q^2}\, ~~\text{and}~~ z\equiv 
        \frac{t}{q^2}
\end{align}
which satisfies the constraint $x+y+z=1$. All the results are presented in terms of Multiple Polylogarithms (MPLs)~\cite{goncharov1998multiple,goncharov2001multiple} containing these parameters. In the next section, we define the form factors in terms of matrix elements.


\subsection{Two- and three-point form factors}
\label{sec:FF}

In perturbation theory, any scattering amplitude can be expressed as expansion in powers of coupling constant 
\begin{align}
\label{eq:mat-exp}
    |{\cal M} \rangle^i_{\{\cal C\}} =& a^{\kappa}\sum_{n=0}^{\infty} a^n |{\cal M}^{(n)}\rangle^i_{\{{\cal C}\}}
\end{align}
where the quantity $|{\cal M}^{(n)}\rangle^i_{\{{\cal C}\}}$ represents the $n$-th loop amplitude of the scattering process depicted in \eqref{eq:process}. The expansion parameter, $a$, is the 't Hooft coupling~\cite{Bern:2005iz} given by
\begin{align}
    a \equiv \frac{g^2 N}{(4\pi)^2} (4 \pi e^{-\gamma_E})^{\epsilon}
\end{align}
with the Euler constant $\gamma_E\approx 0.5772$. $N$ is the quadratic Casimir of SU(N) group in adjoint representation. $\kappa$ in \eqref{eq:mat-exp} corresponds to the power of coupling constant of the leading order amplitude. Form factors are constructed out of the transition matrix elements through
\begin{align}
\label{eq:FF}
    {\cal F}^i_{\{{\cal C}\}} = 1+\sum_{n=1}^{\infty} a^n {\cal F}^{(n),i}_{\{{\cal C}\}} \equiv 1+\sum_{n=1}^{\infty} a^n \left.\frac{\langle {\cal M}^{(0)}|{\cal M}^{(n)} \rangle }{\langle {\cal M}^{(0)}|{\cal M}^{(0)} \rangle}\right\vert^{i}_{\{{\cal C}\}}\,.
\end{align}
The goal of this article is to calculate the three-point FFs to two-loops i.e. ${\cal F}^{(1),i}_{{\cal C}_1{\cal C}_2{\cal C}_3}$, ${\cal F}^{(2),i}_{{\cal C}_1{\cal C}_2{\cal C}_3}$ for the processes in \eqref{eq:procS2}, \eqref{eq:procS}, \eqref{eq:procT}, and two-point FFs up to three-loops i.e. ${\cal F}^{(1),1}_{{\lambda} {\bar \lambda}}$, ${\cal F}^{(2),1}_{{\lambda} {\bar \lambda}}$, ${\cal F}^{(3),1}_{{\lambda} {\bar \lambda}}$ for the processes in \eqref{eq:procO1-2pt}. In the next section, we discuss the methodology of the computations.

\subsection{Methodology of the computation}
\label{sec:calc}
In spite of the presence of many modern techniques bases on unitarity, Feynman diagrammatic approach of evaluating loop amplitudes remains a potential candidate which is employed in this article. We begin by generating the Feynman diagrams of a process using \qgraf~\cite{Nogueira:1991ex}. Because of the presence of Majorana fermions which destroy the flow of fermionic current, a special care needs to be taken. We use a code\footnote{We thank Satyajit Seth for the Python code.} based on \python~ to rectify it which is further processed through a series of in-house \form~\cite{Vermaseren:2000nd} routines in order to apply the Feynman rules, perform Dirac, Lorentz and SU(N) color algebras. To ensure the inclusion of only physical states of the gauge fields, we take into account the ghost loops for internal states and the polarisation sum of the external gluons is performed in axial gauge. For the tensorial interaction, the polarisation sum in $d_{\epsilon}\equiv 4+\epsilon$ dimensions reads~\cite{Mathews:2004xp}
\begin{align}
\label{eq:spin2-pol-sum}
 B^{\mu \nu \rho \sigma}(q) =& \left( g^{\mu\rho} - \frac{\, \, q^{\mu} q^{\rho} }{q.q} \right)  \left(g^{\nu\sigma} - \frac{\, \,q^{\nu} q^{\sigma} }{q.q} \right) 
                           + \left( g^{\mu\sigma} - \frac{\, \,q^{\mu} q^{\sigma} }{q.q} \right)  \left(g^{\nu\rho} - \frac{\, \,q^{\nu} q^{\rho} }{q.q} \right) 
\nonumber \\
& -\, \, \frac{2}{d_{\epsilon}-1} \left( g^{\mu\nu} - \frac{\, \,q^{\mu} q^{\nu} }{q.q} \right) \left( g^{\rho\sigma} - \frac{\, \,q^{\rho} q^{\sigma} }{q.q} \right)\,.
\end{align}
Besides the complicated terms in tensorial operator ${\cal O}^{3}_{\mu\nu}$ in \eqref{eq:op-def-EM}, the presence of several terms in polarisation sum in \eqref{eq:spin2-pol-sum} makes the calculation worse by surging the size of the expression substantially. After performing the Dirac and Lorentz algebras in $d_{\epsilon}$ dimensions and contracting all the indices, the Feynman diagrams manifest as scalar integrals in $d_{\epsilon}$ dimensions. Using the liberty of transforming loop momentum, with the help of \reduze~\cite{vonManteuffel:2012np,Studerus:2009ye}, we categorise all the scalar integrals into 3- and 6-integral families~\cite{Ahmed:2014gla,Ahmed:2014pka} at one- and two-loop, respectively. All the scalar integrals are later related to a much smaller set of master integrals (MIs) through integration-by-parts (IBP)~\cite{Tkachov:1981wb,Chetyrkin:1981qh} and Lorentz invariant (LI)~\cite{Gehrmann:1999as} identities. We use \litered~\cite{Lee:2008tj,Lee:2012cn} along with \mint~\cite{Lee:2013hzt} to perform the integral reduction which ultimately relates all the scalar integrals to 7 and 89 MIs at one- and two-loop, respectively whose results are available in~\cite{Gehrmann:2000zt,Gehrmann:2001ck}. The results are expressed in terms of harmonic polylogarithms (HPLs)~\cite{Gehrmann:1999as} and two-dimensional HPLs (2dHPLs)~\cite{Gehrmann:2000zt,Gehrmann:2001ck,Gehrmann:2001jv}. Using the results of the MIs, we get all the FFs to two-loops which are presented explicitly in the appendix~\ref{app:results}. In the next section, we discuss the regularisation procedure, operator renormalisation and infrared factorisation.

\section{Regularisation prescription and renormalisations}
\label{sec:reg}

The ${\cal N}=4$ SYM is one of the few ultraviolet finite quantum field theories in 4 space-time dimensions. The on-shell amplitudes and FFs of protected operators are guaranteed to be UV finite. However, unprotected operators do develop divergences arising from the UV sector owing to the short distance effects which gets reflected through the presence of non-zero UV anomalous dimensions. Besides, due to the presence of massless vertices in the ${\cal N}=4$ Lagrangian, the theory is not soft and collinear (IR) finite. Consequently, the on-shell amplitudes as well as FFs involve IR divergences. To identify these divergences, we need to regularise the theory following appropriate prescription. In this article, we adopt  supersymmetry preserving modified dimensional reduction (${\overline{\rm DR}}$) scheme~\cite{Siegel:1979wq,Capper:1979ns} where the bosonic and fermionic degrees of freedom (DOF) are maintained to be equal throughout the calculation. In order to achieve that the number of scalar and pseudo-scalar generations are changed from $n_g=3$ to $n_{g,\epsilon}=3-\epsilon/2$ in $d_{\epsilon}$ dimensions. The Majorana fermions are remained to have 4 generations. Together with $(2+\epsilon)$ DOF of gauge fields in $d_{\epsilon}$ dimensions, the total bosonic DOF equals 8 which is same as that of fermions, consequently the SUSY remains intact. The Dirac algebra
and the traces of $\alpha,\beta$ matrices, being dependent on the generation number of the scalar and pseudo-scalar fields, are performed in $d_{\epsilon}$ dimensions which obey
\begin{align}
    \alpha^i\alpha^i=\beta^i\beta^i=(-3+\frac{\epsilon}{2})\mathbb{I}_{4{\rm x}4},~~ \alpha^i\alpha^j\alpha^i=\alpha^j(1-\frac{\epsilon}{2})\mathbb{I}_{4{\rm x}4},~~
    \beta^i\beta^j\beta^i=\beta^j(1-\frac{\epsilon}{2})\mathbb{I}_{4{\rm x}4}\,.
\end{align}
The $\mathbb{I}_{4{\rm x}4}$ denotes 4x4 identity matrix. In the next section, we discuss about the UV renormalisation.

\subsection{UV renormalisation}
\label{ss:uv}
The on-shell amplitudes in ${\cal N}=4$ SYM does not suffer from UV singularities owing to the vanishing of $\beta$-functions to all orders in perturbation theory. However, the FFs of an unprotected composite operator may develop UV divergences which can be manifested by the presence of non-zero UV anomalous dimensions in the FFs beyond leading order. Thus, any FF involving an unprotected operator needs
to go through UV renormalisation which is performed by multiplying an overall operator renormalisation constant, $Z^{i}(a(\mu^2),\epsilon)$, defined through
\begin{align}
\label{eq:Zi-RGE}
    \frac{d}{d\log \mu^2} \log Z^i(a(\mu^2),\epsilon)=\gamma^i=\sum_{k=1}^{\infty} a^k(\mu^2)\gamma^i_k\,.
\end{align}
The UV anomalous dimensions of the operator ${\hat {\cal O}}^i$ are denoted through $\gamma^i$. This is a property inherently associated to the operator itself, it does not depend upon the nature of on-shell states. In particular, the bare operator, ${\hat {\cal O}^i}$, gets renormalised through
\begin{align}
\label{eq:oper-renorm}
    {\cal O}^i=Z^i {\hat {\cal O}}^i\,.
\end{align}
The quantities $Z^{i}(a(\mu^2),\epsilon)$ can be determined by exploiting the universal infrared structures of the two- and three-point FFs and solving the underlying renormalisation group equation satisfied by $Z^{i}(a(\mu^2),\epsilon)$~\cite{Ahmed:2015qpa,Banerjee:2018yrn}. In section.~\ref{ss:mixing}, we discuss the computation of operator renormalisation constants for the operators ${\cal O}^1$ and ${\cal O}^2$. Being conserved to all orders, the $Z^3$ is identically equal to 1. The scale $\mu$ appearing in \eqref{eq:Zi-RGE} is introduced through
\begin{align}
    {\hat a}=a(\mu^2)  \left(\frac{\mu_0^2}{\mu^2} \right)^{\frac{\epsilon}{2}}
\end{align}
where, $\hat a$ is the coupling constant appeared in the regularised Lagrangian in $d_{\epsilon}$ dimensions and $\mu_0$ is the scale introduced to make $\hat a$ dimensionless. 

The UV renormalised matrix elements can be expanded in powers of bare as well as renormalised coupling through
\begin{align}
\label{eq:un-ren}
    |{\cal M} \rangle^{i}_{\{\cal C\}} &= Z^{i}({\hat a},\epsilon)  \sum_{n=0}^{\infty} \left( \frac{\hat a}{\mu_0^{\epsilon}} \right)^{n+\sigma\frac{1}{2}}|{\hat {\cal M}}^{(n)}\rangle^{i}_{\{\cal C\}}\nonumber\\
    &= 
    Z^{i}({a}(\mu^2),\epsilon)  \sum_{n=0}^{\infty} \left( \frac{a(\mu^2)}{\mu^{\epsilon}} \right)^{n+\sigma\frac{1}{2}}|{ {\cal M}}^{(n)}\rangle^{i}_{\{\cal C\}}
\end{align}
where the hat signifies the unrenormalised quantities. The quantity $\sigma=1$ for three-point and $\sigma=0$ for two-point FFs. This, in turn, leads us to the expressions of renormalised matrix elements in terms of bare ones order by order in coupling constant.
%
Employing these relations in \eqref{eq:FF}, we determine the UV renormalised FFs at 1-, 2- and 3-loop which are computed in this article. Now we turn our focus towards the infrared factorisation of the FFs.

\subsection{Infrared factorisation}
\label{ss:ir}
Being a massless theory, the UV renormalised FFs exhibit infrared divergences (IR) from soft and collinear configurations. These appear as poles in dimensional regularisation parameter $\epsilon$, whose universal nature was first demonstrated in~\cite{Catani:1998bh}. In~\cite{Sterman:2002qn}, a detailed derivation was presented by exploiting the factorisation and resummation properties of scattering amplitudes which was later generalised to all orders in~\cite{Becher:2009cu,Gardi:2009qi}. The IR divergences can be factored out from UV renormalised quantities through the subtraction operators $\mathbf{I}^{(n)}_{\{\cal C\}}$ which up to two-loops read
\begin{align}
\label{eq:mat-cat}
    |{\cal M}^{(1)}\rangle^{i}_{\{\cal C\}} &= 2 {\mathbf{I}}^{(1)}_{\{\cal C\}}(\epsilon) |{\cal M}^{(0)}\rangle^{i}_{\{\cal C\}} + |{\cal M}^{(1)}_{\rm fin}\rangle^{i}_{\{\cal C\}}\,,\nonumber\\
    |{\cal M}^{(2)}\rangle^{i}_{\{\cal C\}} &= 4 {\mathbf{I}}^{(2)}_{\{\cal C\}}(\epsilon) |{\cal M}^{(0)}\rangle^{i}_{\{\cal C\}} + 2 {\mathbf{I}}^{(1)}_{\{\cal C\}}(\epsilon) |{\cal M}^{(1)}\rangle^{i}_{\{\cal C\}} + |{\cal M}^{(2)}_{\rm fin}\rangle^{i}_{\{\cal C\}}\,.
\end{align}
The subtraction operators are independent of the nature of operator insertion, it solely depends on the SU(N) coloured particles. In case of ${\cal N}=4$ SYM where all the fundamental particles transform under the adjoint representation of SU(N) gauge group and coefficients of $\beta$-function vanishes to all orders, these take the form
\begin{align}
\label{eq:sub}
    &{\mathbf{I}}^{(1)}_{{{\cal C}_1{\cal C}_2}}(\epsilon)=-\frac{e^{-\frac{\epsilon}{2}\gamma_E}}{\Gamma(1+\frac{\epsilon}{2})} \left(\frac{4}{\epsilon^2}\right) \left( -\frac{s}{\mu^2} \right)^{\frac{\epsilon}{2}}   \,,\nonumber\\
    &{\mathbf{I}}^{(1)}_{{{\cal C}_1{\cal C}_2{\cal C}_3}}(\epsilon)=-\frac{e^{-\frac{\epsilon}{2}\gamma_E}}{\Gamma(1+\frac{\epsilon}{2})} \left(\frac{2}{\epsilon^2}\right) \Bigg\{ \left( -\frac{s}{\mu^2} \right)^{\frac{\epsilon}{2}} +\left( -\frac{t}{\mu^2} \right)^{\frac{\epsilon}{2}} +\left( -\frac{u}{\mu^2} \right)^{\frac{\epsilon}{2}} \Bigg\} \,,\nonumber\\
    &{\mathbf{I}}^{(2)}_{{{\cal C}_1{\cal C}_2}}(\epsilon)=-\frac{1}{2} \left(\mathbf{I}^{(1)}_{{{\cal C}_1{\cal C}_2}}(\epsilon)\right)^2 - e^{\frac{\epsilon}{2}\gamma_E}\frac{ \Gamma(1+\epsilon)}{\Gamma(1+\frac{\epsilon}{2})} \zeta_2~ \mathbf{I^{(1)}}_{{{\cal C}_1{\cal C}_2}}(2\epsilon) + \frac{1}{\epsilon} \mathbf{H}^{(2)}_{{{\cal C}_1{\cal C}_2}}(\epsilon)\,,\nonumber\\
    &{\mathbf{I}}^{(2)}_{{{\cal C}_1{\cal C}_2{\cal C}_3}}(\epsilon)=-\frac{1}{2} \left(\mathbf{I}^{(1)}_{{{\cal C}_1{\cal C}_2{\cal C}_3}}(\epsilon)\right)^2 - e^{\frac{\epsilon}{2}\gamma_E}\frac{ \Gamma(1+\epsilon)}{\Gamma(1+\frac{\epsilon}{2})} \zeta_2~ \mathbf{I^{(1)}}_{{{\cal C}_1{\cal C}_2{\cal C}_3}}(2\epsilon) + \frac{1}{\epsilon} \mathbf{H}^{(2)}_{{{\cal C}_1{\cal C}_2{\cal C}_3}}(\epsilon)
\end{align}
with~\cite{Becher:2009cu} 
\begin{align}
    &\mathbf{H}^{(2)}_{{{\cal C}_1{\cal C}_2}}(\epsilon) = -\frac{1}{4}\zeta_3\,,\nonumber\\
    &\mathbf{H}^{(2)}_{{{\cal C}_1{\cal C}_2{\cal C}_3}}(\epsilon) = -\frac{3}{4}\zeta_3.
\end{align}
The arbitrariness of the finite parts of the subtraction operators define various schemes in which finite part of amplitude is computed. Upon translating the infrared factorisation from amplitudes to the UV renormalised FFs depicted in \eqref{eq:FF}, we get exactly same relations as \eqref{eq:mat-cat}:
\begin{align}
\label{eq:fffin}
    &{\cal F}^{(1),i}_{\{\cal C\}}(\epsilon)=2 {\mathbf{I}}^{(1)}_{\{\cal C\}}(\epsilon) + {\cal F}^{(1),i}_{\{{\cal C}\},\rm fin}\,,\nonumber\\
    &{\cal F}^{(2),i}_{\{\cal C\}}(\epsilon)=4 {\mathbf{I}}^{(2)}_{\{\cal C\}}(\epsilon) + 2 {\mathbf{I}}^{(1)}_{\{\cal C\}}(\epsilon) {\cal F}^{(1),i}_{\{\cal C\}}(\epsilon) + {\cal F}_{\{\cal C\},\rm fin}^{(2),i}\,.
\end{align}
The infrared structure of the Sudakov FF at three-loop can be obtained by solving the renormalisation group equation satisfied by the Sudakov FF~\cite{Sudakov:1954sw,Mueller:1979ih,Collins:1980ih,Sen:1981sd}. Using the solution depicted in refs.~\cite{Ravindran:2005vv,Ravindran:2006cg,Moch:2005tm}, we obtain the universal infrared structures (See ref.~\cite{Ahmed:2016vgl}). Computation of the finite quantities ${\cal F}^{(1),i}_{\{\cal C\},\rm fin}(\epsilon)$, ${\cal F}^{(2),i}_{\{\cal C\},\rm fin}(\epsilon)$ and ${\cal F}^{(3),i}_{\{\cal C\},\rm fin}(\epsilon)$ is the primary objective of this article. The finite parts of the FFs appearing on the right side of the \eqref{eq:fffin} can be organised according to their transcendental weights as
\begin{align}
\label{eq:FF-expand-tr-weight}
    {\cal F}_{\{\cal C\},\rm fin}^{(n),i} = 
    \sum_{k=0}^{2n} {\cal F}_{\{\cal C\},\rm fin}^{(n),i,\tau(k)}\,.
\end{align}
The explicit results of the FFs in terms of MPLs are presented in the appendix~\ref{app:results} by setting $\mu^2=-q^2$. In the following section, we discuss the operator mixing in the context of operator renormalisation.

\subsection{Operator mixing and dilatation operator to three-loops in SU($2|3$) sector}
\label{ss:mixing}
The operators ${\cal O}^i$ for $i=1,2$ are composed of bosonic and fermionic parts which we denote as ${\cal O}^i_B$ and ${\cal O}^i_F$, respectively and those are defined through
\begin{align}
\label{eq:O-OB-OF}
    &{\cal O}^1_F={\bar \lambda}^a_m  {\lambda}^a_m\,,\qquad\qquad {\cal O}^1_B=4 g\,f^{abc} \varepsilon_{ijk} \phi^a_i \phi^b_j \phi^c_k\,,\nonumber\\
    &{\cal O}^2_F=-\frac{g}{4} {\cal O}^1_F\,,\qquad\qquad {\cal O}^2_B=\frac{{\cal O}^1_B}{4g}\,.
\end{align}
These constituents operators $\mathcal{O}^i_B$ and $\mathcal{O}^i_F$ mix among themselves under UV renormalisation through
\begin{align}
  \label{eq:OpMat}
  \mathcal{O}_{\Sigma}^{i} &= {\cal Z}_{\Sigma \Lambda}^{i}  \hat{\mathcal{O}}_{\Lambda}^{i}\,, 
\end{align}
with the mixing matrix
\begin{align}
  \label{eq:ZMat}
    {\cal Z}^{i} \equiv
    \begin{bmatrix}
    {\cal Z}_{BB}^{i}  & {\cal Z}_{BF}^{i} \\
    {\cal Z}_{FB}^{i}  & {\cal Z}_{FF}^{i}
    \end{bmatrix}\,,
\end{align}
and $\Sigma, \Lambda\in\{B,F\}$. The bare operator is denoted by ${\hat{\cal O}}^i$. The operator renormalisation constant is controlled by the corresponding anomalous dimensions, $\gamma^i_{lm}$, through renormalisation group equation (RGE)
\begin{align}
\label{eq:RG}
    \mu^2 \frac{d}{d\mu^2} {{\cal Z}}^i_{lm}= \gamma^i_{lk} {{\cal Z}}^i_{km}\,.
\end{align}
The general solution of the RGE up to ${\cal O}(a^3)$ is obtained as (see the appendix of ref.~\cite{Ahmed:2017rwl} for general method of solving RGE)
\begin{align}
  \label{eq:ZCoupSoln}
  {\cal Z}_{lm}^{i}  &= \delta_{lm}^{}
           + {a} \Bigg[ \frac{2}{\epsilon}
           \gamma_{lm,1}^{i} \Bigg] 
           + {a}^{2} \Bigg[
           \frac{1}{\epsilon^{2}} \Bigg\{ 
            2 \gamma_{lk,1}^{i} \gamma_{km,1}^{i}  \Bigg\} + \frac{1}{\epsilon} \Bigg\{ \gamma_{lm,2}^{i}\Bigg\}
           \Bigg]\nonumber\\& + a^{3} \Bigg[ \frac{1}{\epsilon^{3}} \Bigg\{
            \frac{4}{3} \gamma_{lk,1}^{i} \gamma_{kn,1}^{i}
           \gamma_{nm,1}^{i} \Bigg\} + \frac{1}{\epsilon^{2}} \Bigg\{ 
           \frac{2}{3}
           \gamma_{lk,1}^{i} \gamma_{km,2}^{i}
           \nonumber\\
           &+ \frac{4}{3}
           \gamma_{lk,2}^{i} \gamma_{km,1}^{i} \Bigg\} + \frac{1}{\epsilon}
           \Bigg\{ \frac{2}{3} \gamma_{lm,3}^{i} \Bigg\} \Bigg]\,,
\end{align}
where $\gamma_{lm}^{i}$ are expanded in powers of $a$ as 
\begin{align}
\label{eq:gamma-expand}
    \gamma_{lm}^i=\sum_{n=1}^{\infty} a^n \gamma_{lm,n}^{i}\,.
\end{align}
Using the results of the FFs of the renormalised operators, computed in this article, and by demanding the UV finiteness and universality of the infrared singularities, we have computed the operator renormalisation constants or equivalently, the underlying anomalous dimensions. In order to achieve this, we have employed the methodology prescribed in refs.~\cite{Ahmed:2015qpa,Ahmed:2016qjf,Banerjee:2018yrn}. The one-loop results are found to be
\begin{align}
\label{eq:res-gamma-1loop}
 &\gamma_{FF,1}^{1}=0\,, \quad  \gamma_{BF,1}^{1}=0\,, \quad \,  \gamma_{BB,1}^{1}=-6\,, \quad   \gamma_{FB,1}^{1}=6\,.\nonumber\\
 &\gamma_{FF,1}^{2}=0\,, \quad  \gamma_{BF,1}^{2}= -6\,, \quad \,   \gamma_{BB,1}^{2}= -6 \,, \quad \gamma_{FB,1}^{2}=0\,.  
 \end{align}
From the aforementioned results of the anomalous dimensions, we observe that the diagonal elements, namely, $\gamma_{BB,1}^{i}$ and $\gamma_{FF,1}^{i}$ for the two operators are identical. At two-loop, the results of the anomalous dimensions are obtained as
\begin{align}
\label{eq:res-gamma-2loop}
 &\gamma_{FF,2}^{1}= -12 \,, \quad  \gamma_{BF,2}^{1}=12\,, \quad \,  \gamma_{BB,2}^{1}=36\,, \quad   \gamma_{FB,2}^{1}=-48\,,\nonumber\\
 &\qquad\qquad\qquad\qquad\qquad\qquad\quad~~\gamma_{BB,2}^{2}= 36\,, \quad   \gamma_{FB,2}^{2}= -12\,.
\end{align}
In order to determine the remaining two components of the anomalous dimensions corresponding to the operator ${\cal O}^2$, the computations of three-point three-loop or two-point four-loop FFs are required, which are beyond the scope of this article. In addition to determining these quantities to two-loops, by exploiting the three-loop FFs of the operator ${\cal O}^1$, we have also determined some of the elements of the three-loop anomalous dimension matrix, which read~\footnote{The three-loop anomalous dimensions are dependent on $\omega$ which is defined through $\varepsilon_{ijk} \varepsilon_{ijk}=6+\omega \epsilon$. In ${\overline{\rm DR}}$ regularisation scheme, $\omega$ equals $-\frac{11}{2}$ which is the coefficient of $\epsilon$ in the expansion of $n_{g,\epsilon}=(3-\frac{\epsilon}{2})!$. $n_{g,\epsilon}$ is the number of generations of scalar/pseudo-scalar in ${\overline{\rm DR}}$, as mentioned in~\ref{sec:reg}.}
\begin{align}
\label{eq:res-gamma-3loop}
 \gamma_{FF,3}^{1}=  \,18, \quad  \gamma_{BF,3}^{1}=18\,.
\end{align}
Our findings are consistent with the partially available two-loop results computed in ref.~\cite{Brandhuber:2016fni}. The components $\gamma_{FB,2}^1$ at two-loop, and $\gamma_{FF,3}^1$, $\gamma_{BF,3}^1$ at three-loop are new results computed for the first time in this article. The anomalous dimensions of ${\cal O}^2$ are found to be identical to the Konishi supermultiplet, in accordance with the expectations~\cite{Eden:2005ve,Brandhuber:2016fni}. For readers' convenience, below we present the mixing matrices explicitly, which read
\begin{align}
\label{eq:mixing-matrix-results}
    	&{\cal Z}^1=
	\begin{bmatrix}
	1 -a \frac{12}{\epsilon} + a^2 \left( \frac{72}{\epsilon^2} + \frac{36}{\epsilon} \right) + a^3 \left( - \frac{288}{\epsilon^3}- \frac{336}{\epsilon^2}  + \frac{2 \gamma_{BB,3}}{3\epsilon}\right) & ~~a^2 \frac{12}{\epsilon} + a^3 \left( -\frac{48}{\epsilon^2} +\frac{12}{\epsilon}\right) \\
	a \frac{12}{\epsilon} + a^2 \left( -\frac{72}{\epsilon^2} - \frac{48}{\epsilon} \right) + a^3 \left( \frac{288}{\epsilon^3} + \frac{432}{\epsilon^2} + \frac{2 \gamma_{FB,3}}{3\epsilon} \right) & ~~1-a^2 \frac{12}{\epsilon} + a^3 \left( \frac{48}{\epsilon^2} +\frac{12}{\epsilon}\right)
	\end{bmatrix}\,,\nonumber\\
	&{\cal Z}^2=
 	\begin{bmatrix}
 	 1-a \frac{12}{\epsilon} + a^2 \left( \frac{72}{\epsilon^2} + \frac{36}{\epsilon}\right) &~~- a \frac{12}{\epsilon} + a^2 \left(\frac{72}{\epsilon^2} + \frac{\gamma_{BF,2}^2}{\epsilon} \right) \\
 	 - a^2 \frac{12}{\epsilon}  & ~~1+a^2 \frac{\gamma_{FF,2}^2}{\epsilon}
	\end{bmatrix}\,.
\end{align}

The quantum corrections to the dilatation operators corresponding to the operators ${\cal O}^i_B$ and ${\cal O}^i_F$ in the SU($2|3$) sector~\cite{Beisert:2003ys,Brandhuber:2016fni} are defined as
\begin{align}
\label{eq:dilatation-defn}
	\delta {\cal D}^i \equiv \lim_{\epsilon \rightarrow 0} \left[ - \mu^2 
	\frac{\partial}{\partial\mu^2}  \log({\cal Z}^i)\right]\,.
\end{align}
This is essentially the anomalous dimension matrix with a difference of an overall factor of -1. For readers' convenience, we also present the results of the $\log({\cal Z}^i)$ below
\begin{align}
\label{eq:logZ-O1}
	&\log({\cal Z}^1)=
	\begin{bmatrix}
	 -a \frac{12}{\epsilon} + a^2 \frac{36}{\epsilon} + a^3 \left( \frac{24}{\epsilon^2} + \frac{2 \gamma_{BB,3}}{3\epsilon}\right) & ~~a^2 \frac{12}{\epsilon} + a^3 \left( \frac{24}{\epsilon^2} +\frac{12}{\epsilon}\right) \\
	a \frac{12}{\epsilon} - a^2 \frac{48}{\epsilon} + a^3 \frac{2 \gamma_{FB,3}}{3\epsilon} & ~~-a^2 \frac{12}{\epsilon} + a^3 \left(- \frac{24}{\epsilon^2} +\frac{12}{\epsilon}\right)
	\end{bmatrix}\,,\nonumber\\
	&\log({\cal Z}^2)=
 	\begin{bmatrix}
 	 -a \frac{12}{\epsilon} + a^2 \frac{36}{\epsilon}  &~~- a \frac{12}{\epsilon} + a^2 \frac{\gamma_{BF,2}^2}{\epsilon} \\
 	 - a^2 \frac{12}{\epsilon}  & ~~a^2 \frac{\gamma_{FF,2}^2}{\epsilon}
	\end{bmatrix}\,.
\end{align}
The remaining components need computations which are beyond the scope of this paper. The results of the $\log({\cal Z}^1)$ which were computed partially to two-loops level in ref.~\cite{Brandhuber:2016fni} are consistent with our findings. 

Using the results of the mixing matrix, we can get the renormalised operator defined in \eqref{eq:oper-renorm} as
\begin{align}
\label{eq:oper-renorm-ITO-mixing}
    {\cal O}^i = \left( {\cal Z}_{BB}^i + {\cal Z}_{FB}^i \right) {\hat {\cal O}}_B
    + \left( {\cal Z}_{BF}^i + {\cal Z}_{FF}^i \right) {\hat {\cal O}}_F
\end{align}
where, the ${\cal Z}^i_{lm}$ can be directly read from \eqref{eq:mixing-matrix-results}. Upon substituting the results of the mixing matrix from \eqref{eq:mixing-matrix-results}, we have seen that both the operators ${\cal O}^1$ and ${\cal O}^2$ have non-zero anomalous dimensions. Moreover, the anomalous dimension of the operator ${\cal O}^2$ turns out to be identical to that of Konishi supermultiplet, in accordance with the expectations~\cite{Eden:2005ve,Brandhuber:2016fni}.

\section{Finite remainders at two-loop}
\label{sec:FinRem}
For maximally helicity violating (MHV) amplitude in planar ${\cal N}=4$ SYM, the helicity blind quantity which is obtained by factoring out the tree level amplitude was conjectured to exponentiate in terms of one-loop counter part along with the universal cusp and collinear anomalous dimensions. This is known as BDS/ABDK ansatz~\cite{Anastasiou:2003kj,Bern:2005iz}. However, this ansatz is found to break down for two-loop six-point amplitudes~\cite{Bern:2008ap,Drummond:2008aq}. 
\begin{figure}[htb]
\begin{center}
\includegraphics[scale=0.7]{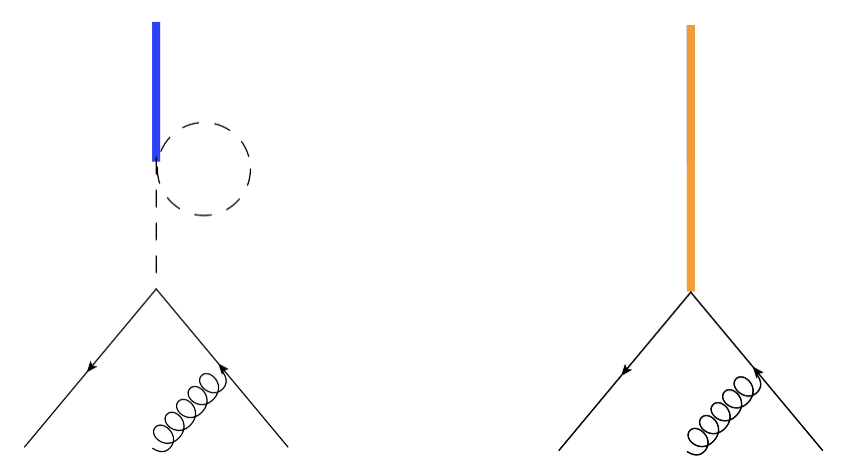}
\caption{Sample of Feynman diagrams contributing to the leading order process $J^2 \rightarrow g\lambda\bar\lambda$. The blue and orange colour denote the operator insertions of ${\cal O}^2_B$ and ${\cal O}^2_F$ which are defined in \eqref{eq:O-OB-OF}. The solid line with containing arrow, curly and dash lines represent the fermion, gluon and scalar particles. (The diagram on the left hand side is zero in dimensional regularisation. We draw these just for the purpose of illustration.)}
\label{dia:O2gll}
\end{center}
\end{figure}
Though the IR divergent parts were found to be consistent with the prediction of the ansatz, the finite part was not. Hence, to capture the amount of deviation from the prediction of BDS/ABDK ansatz, a quantity named finite remainder~\cite{Bern:2008ap,Drummond:2008aq} was introduced which is a finite function of dual conformal cross ratios. 

Following this line of thought and inspired by the exponentiation of the IR divergences of FFs, the finite remainder function for FFs at two-loop was introduced in ref.~\cite{Brandhuber:2012vm} as
\begin{align}
\label{eq:fin-rem}
\mathcal{R}^{(2),i}_{\{ \cal C\}} \equiv \mathcal{F}^{(2),i}_{\{ \cal C\}}(\epsilon)-\frac{1}{2}\left(\mathcal{F}^{(1),i}_{\{ \cal C\}}(\epsilon)\right)^2-f^{(2)}(\epsilon)\mathcal{F}^{(1),i}_{\{ \cal C\}}(2\epsilon)-C^{(2)} + {\cal O}(\epsilon)
\end{align}
where $f^{(2)}(\epsilon) \equiv \sum_{j=0}^2 \epsilon^j f_j^{(2)}$. Writing a two-loop FF in terms a quantity dictated by BDS/ABDK ansatz plus a finite remainder provides an efficient way of organising the result - the ansatz part captures the IR divergences whereas the remainder encapsulates the finite part in 4-dimensions. Due to the nature of IR divergences of FFs, the $1/\epsilon^4$ and $1/\epsilon^3$ poles cancel between first two terms in \eqref{eq:fin-rem}, By demanding the finiteness of remainder i.e. the vanishing of the remaining $1/\epsilon^2$ and $1/\epsilon$ poles, one gets
\begin{align}
    f_0^{(2)}=-2\zeta_2\,, \quad\quad f_1^{(2)}=\zeta_3.
\end{align}
The other coefficient $f_2^{(2)}$ and $C^{(2)}$ can not be fixed uniquely. To maintain a close resemblance with the corresponding quantity of MHV amplitudes, 
the $f_2^{(2)}$ is assigned a value according to
\begin{align}
    f_2^{(2)}=-\frac{1}{2} \zeta_4\,.
\end{align}
With the help of Sudakov FF, the remaining constant is found to be~\cite{Brandhuber:2012vm}
\begin{align}
     C^{(2)}=4\zeta_4\,.
\end{align}
$f^{(2)}(\epsilon)$ and $C^{(2)}$ are independent of the nature of operator and number of external legs. The remainder $\mathcal{R}^{(2),i}_{\{ \cal C\}}$ and finite part of the ${\cal F}_{\{ \cal C\},\rm fin}^{(2),i}$ can be related to each other which can be found in~\cite{Banerjee:2016kri}. In this article, we compute the finite remainders for all the processes and operators which are presented explicitly in the appendix~\ref{app:results} after setting $\mu^2=-q^2$ and organising those according to their transcendental weights as
\begin{align}
\label{eq:FR-expand-tr-weight}
\mathcal{R}^{(2),i}_{\{ \cal C\}}
= \sum_{k=0}^{4} \mathcal{R}^{(2),i,\tau(k)}_{\{ \cal C\}}\,.
\end{align}
We do not define the finite remainder at two-loop for the processes $J^2 \rightarrow g\lambda\bar\lambda$ and $J^2 \rightarrow \phi\lambda\bar\lambda$ due to the involvement of one-loop diagrams arising from ${\cal O}^2_F$. This can be understood by looking at the leading order contributions, shown in figure~\ref{dia:O2gll}, to these processes which are loop induced through ${\cal O}^2_B$. In the next section, we analyse our results in view of transcendetality weight.

\section{Principle of uniform transcendentality}
\label{sec:lt}
The principle of uniform transcendentality (PUT) is a remarkable feature based on numerous observations~\cite{Kotikov:2002ab,Kotikov:2004er,Bern:2006ew,Drummond:2007cf,Naculich:2008ys,Bork:2010wf,Gehrmann:2011xn,Brandhuber:2012vm,Eden:2012rr,Drummond:2013nda,Basso:2015eqa,Goncalves:2016vir,Banerjee:2018yrn}, albeit unproven, which states that certain kinds of scattering amplitudes and form factors in ${\cal N}=4$ can be expressed purely in terms of functions having UT. In other words, the L-th loop result is found to contain only polylogarithmic functions of degree 2L. 
It is conjectured in ref.~\cite{Loebbert:2015ova} that at two-loop level the HT weight parts of every two-point minimal FF (number of fields present in the operator is same as number of external on-shell states) are identical and those are same as that of half-BPS operator belonging to the stress-energy supermultiplet~\cite{vanNeerven:1985ja}. In ref.~\cite{Brandhuber:2014ica}, the minimal FFs of long BPS operators (more than two fields) are computed to two-loops, and the corresponding HT piece is found to be a universal in all FFs with long, unprotected operators~\cite{Loebbert:2015ova,Brandhuber:2016fni,Brandhuber:2017bkg,Jin:2018fak,Jin:2019ile,Brandhuber:2018kqb}. 
It is also conjectured in~\cite{Loebbert:2015ova}, that the HT terms of two-loop remainder of the three-point FF of every length-two operator should agree with the corresponding half-BPS remainder found in ref.~\cite{Brandhuber:2012vm} which is falsified in ref.~\cite{Banerjee:2016kri} where, for the first time, it is shown that for three-point FF of the Konishi operator (length-two), the HT part depends on the nature of external on-shell states; it fails to match with that of half-BPS if the external on-shell states are $\phi\lambda\lambda$. 
The operator ${\cal O}^1$ and ${\cal O}^2$ in \eqref{eq:op-def} are combination of short (length-two) and long (length-three) unprotected parts. Moreover, the corresponding FFs can neither be regarded as minimal nor non-minimal case. So, this scenario does not belong to the domains for which the aforementioned conjectures are made. This kind of scenario is largely unexplored. 
In this section, we dissect our new results in light of these principles and conjectures, and discuss how far the PUT holds true in present context.

\begin{itemize}
    \item In case of tensorial interaction through operator ${\cal O}^{3}_{\mu\nu}$, for all the processes depicted in \eqref{eq:procT}, the one-loop and two-loop finite parts of the three-point FFs i.e. ${\cal F}_{{\cal C}_1{\cal C}_2{\cal C}_3,\rm fin}^{(1),3}$, ${\cal F}_{{\cal C}_1{\cal C}_2{\cal C}_3,\rm fin}^{(2),3}$ in \eqref{eq:fffin}, are UT. These contain only the HT terms which involve exclusively polylogarithmic functions of transcendentality 2 and 4 at one- and two-loop, respectively. No lower transcendental terms is present. Moreover, the corresponding finite remainder $\mathcal{R}^{(2),3}_{{\cal C}_1{\cal C}_2{\cal C}_3}$ in \eqref{eq:fin-rem} replicates this behaviour which is exactly similar to the corresponding quantity of the half-BPS operator. More specifically, the results in terms of MPLs are obtained as
    \begin{align}
    \label{eq:FF-Tensor-reln-hBPS}
        &{\cal F}^{(n),{3}}_{ggg,\rm fin} = {\cal F}^{(n),{3}}_{g\phi\phi,\rm fin} = {\cal F}^{(n),{3}}_{g\lambda\bar\lambda,\rm fin}={\cal F}^{(n),{3}}_{\phi\lambda\bar\lambda,\rm fin} \equiv {\cal F}^{(n),{3}, \tau(2n)}_{{\cal C}_1{\cal C}_2{\cal C}_3,\rm fin} = 
        {\cal F}^{(n),\text{half-BPS}}_{{\cal C}_1{\cal C}_2{\cal C}_3,\rm fin}\,,\nonumber\\
        &{\mathcal R}^{(2),{3}}_{ggg} = {\mathcal R}^{(2),{3}}_{g\phi\phi} = {\mathcal R}^{(2),{3}}_{g\lambda\bar\lambda}={\mathcal R}^{(2),{3}}_{\phi\lambda\bar\lambda}\equiv{\mathcal R}^{(2),{3}, \tau(4)}_{{\cal C}_1{\cal C}_2{\cal C}_3} = 
        {\mathcal R}^{(2),\text{half-BPS}}_{{\cal C}_1{\cal C}_2{\cal C}_3} \,,
    \end{align}
    where $\tau(i)$ represents terms with transcendentality weight $i$. The above eq. \eqref{eq:FF-Tensor-reln-hBPS} implies that the finite FFs as well as the remainders are independent of the on-shell states. Moreover, the results are exactly same as that of the half-BPS operator~\cite{Brandhuber:2012vm,Banerjee:2016kri} which is presented explicitly in terms of MPLs in appendix~\ref{app:results-hBPS}. In summary, the energy-momentum tensor behaves exactly like the half-BPS operator at three-point FFs level. This is in accordance with our classical expectation: since the stress-tensor is protected and lies in the same multiplet as the half-BPS, an exact SUSY Ward identity would relate these two FFs. 

    \item Through our computations, we find that except the three-point one-loop FF of operator ${\cal O}^1$ with $\phi\phi\phi$ on-shell states, none of the other two- and three-point FFs of ${\cal O}^1$ exhibits UT behaviour. The general observation is that the FFs of SUSY protected operators exhibit UT behaviour. However, there is no clear indication whether the reverse is also true. Through the results depicted in \eqref{eq:FF-O1-ppp-UT}, we see that the FF of an unprotected operator, such as, ${\cal O}^1$ can even be UT. The HT parts of the three-point remainder function and the finite FFs at one- as well as two-loop do match with that of half-BPS~\cite{vanNeerven:1985ja,Gehrmann:2011xn,Brandhuber:2012vm,Banerjee:2016kri}:
    \begin{align}
    \label{eq:FF-O1-reln-hBPS}
        &{\cal F}^{(n),1,\tau(2n)}_{\lambda\bar\lambda,\rm fin} = 
        {\cal F}^{(n),\text{half-BPS}}_{{\cal C}_1{\cal C}_2,\rm fin}\,,\nonumber\\
        &{\cal F}^{(n),1,\tau(2n)}_{g\lambda\bar\lambda,\rm fin}  =
        {\cal F}^{(n),1,\tau(2n)}_{\phi\lambda\bar\lambda,\rm fin} = 
        {\cal F}^{(n),1,\tau(2n)}_{\phi\phi \phi,\rm fin} 
        \equiv{\cal F}^{(n),1, \tau(2n)}_{{\cal C}_1{\cal C}_2{\cal C}_3,\rm fin} = 
        {\cal F}^{(n),\text{half-BPS}}_{{\cal C}_1{\cal C}_2{\cal C}_3,\rm fin}\,,\nonumber\\
        &{\mathcal R}^{(2),1,\tau(4)}_{g\lambda\bar\lambda} = 
        {\mathcal R}^{(2),1,\tau(4)}_{\phi\lambda\bar\lambda}=
         {\mathcal R}^{(2),1,\tau(4)}_{\phi\phi \phi}
        \equiv{\mathcal R}^{(2),1, \tau(4)}_{{\cal C}_1{\cal C}_2{\cal C}_3} = 
        {\mathcal R}^{(2),\text{half-BPS}}_{{\cal C}_1{\cal C}_2{\cal C}_3}\,\\
        {\rm and}&\nonumber\\
        &{\cal F}^{(1),1,\tau(n<2)}_{\phi\phi\phi,\rm fin} =0\label{eq:FF-O1-ppp-UT}\,.
\end{align}
We also see that the HT terms are independent of the external on-shell states. As expected, the lower transcendental weight terms do depend on the nature of external on-shell states. The results of the aforementioned FFs are presented explicitly in appendix~\ref{app:results-O1-2pt} and~\ref{app:results-O1-3pt}.

\item The FFs of the operator $\mathcal{O}^2$ have lower transcendental weight terms, similar to that of ${\cal O}^1$. However, in contrast to the FFs of $\mathcal{O}^1$, the HT terms of this operator depend on the nature of external on-shell states. Moreover, none of the HT terms of the FFs matches with that of half-BPS:
\begin{align}
\label{eq:FF-O2-reln-hBPS}
 &{\cal F}^{(n),2,\tau(2n)}_{{\cal C}_1{\cal C}_2{\cal C}_3,\rm fin} \neq
        {\cal F}^{(n),\text{half-BPS}}_{{\cal C}_1{\cal C}_2{\cal C}_3,\rm fin}\,,\nonumber\\
        &{\mathcal R}^{(2),2,\tau(4)}_{\phi\phi\phi} \neq
        {\cal R}^{(2),\text{half-BPS}}_{{\cal C}_1{\cal C}_2{\cal C}_3,\rm fin} \,.
\end{align}
Similar behaviour in the context of Konishi operator was observed in ref.~\cite{Banerjee:2016kri} for the two-loop three-point FFs with external on-shell states containing $\phi\lambda\bar\lambda$. So, we see that the dependence of HT terms on the nature of external on-shell states is a general feature of the FFs of Konishi primary and its descendants. 

In ref.~\cite{Brandhuber:2014ica}, the minimal FFs of long BPS operators (more than two fields) are computed to two-loops, and the corresponding HT piece is found to be a universal in all FFs with long (more than two-fields), unprotected operators~\cite{Loebbert:2015ova,Brandhuber:2016fni,Brandhuber:2017bkg,Jin:2018fak,Jin:2019ile,Brandhuber:2018kqb}. For the FF of ${\cal O}^2$ with the external on-shell states $\phi\phi\phi$, the one-loop contribution comes only from ${\cal O}^2_B$, the ${\cal O}^2_F$ does not contribute at this loop level. So, this one-loop FF effectively comes from a long, unprotected operator. For this case, we see that the HT part of the one-loop FF does not match with the HT term of long BPS operator, and consequently it violates the conjecture.

The results of the  FFs of operator ${\cal O}^2$ are presented explicitly in appendix~\ref{app:results-O2-3pt}.
\end{itemize}
In the following section, we concentrate only on the highest transcendental weight terms and discuss their profound implications.

\section{Principle of maximal transcendentality and its violation}
\label{sec:PMT}
The principle of maximal transcendentality (PMT) ~\cite{Kotikov:2001sc,Kotikov:2004er,Kotikov:2006ts} establishes a bridge between the results in QCD and those of ${\cal N}=4$ SYM which are comparatively simpler. It states that for certain quantities, the results in ${\cal N}=4$ SYM can be obtained from that in QCD by converting the fermions from fundamental to adjoint representation and then retaining only the HT terms. The complete domain of applicability of this principle is still not clear and under active investigation. In this article, we check the applicability of PMT with respect to the operators considered. 

For all the FFs of operator ${\cal O}^1$ and ${\cal O}^3$, the HT terms of the finite FFs as well as remainder functions coincide with the corresponding results of the half-BPS operator~\cite{vanNeerven:1985ja,Gehrmann:2011xn,Brandhuber:2012vm,Banerjee:2016kri}, as shown in \eqref{eq:FF-O1-reln-hBPS}. However, none of the HT parts of the operator ${\cal O}^2$ matches with that of half-BPS, as shown in \eqref{eq:FF-O2-reln-hBPS}. In refs.~\cite{Brandhuber:2012vm,Banerjee:2017faz,Brandhuber:2017bkg,Jin:2018fak,Jin:2019ile}, it was shown that the three-point FFs of half-BPS are identical to the HT terms of the quark (vector interaction) and gluon (scalar interaction) FFs of QCD~\cite{Gehrmann:2011aa} after converting the fermions from fundamental to adjoint representation. Consequently, with respect to the quark and gluon FFs in QCD, all of the FFs of ${\cal O}^1$ and ${\cal O}^3$, but ${\cal O}^2$, satisfy the PMT.

Now we ask the question, instead of considering the QCD FFs of scalar and vector interactions, if we take the tensorial interaction, does the PMT hold true? Quite remarkably, the answer turns out to be `No': the HT terms of the three-point QCD FFs of EM tensor~\cite{Ahmed:2014gla,Ahmed:2016yox} do not match with the corresponding quantities of ${\cal N}=4$ SYM upon converting the quarks from fundamental to adjoint representation. To demonstrate this, we consider the processes depicted in \eqref{eq:procT} in ${\cal N}=4$ SYM and the corresponding two processes in massless QCD that (a) involves three gluons~\cite{Ahmed:2014gla} and (b) one gluon along with Dirac's fermion-antifermion pair~\cite{Ahmed:2014pka}.

Let's begin with the process (a). We find none of the HT terms of the corresponding FFs of the underlying process in QCD~\cite{Ahmed:2014gla}  match with that of ${\cal N}=4$ SYM in \eqref{eq:FF-Tensor-reln-hBPS} upon converting the fermions from fundamental to adjoint representation through $\{C_F \rightarrow C_A, 2 n_f T_F \rightarrow C_A\}$ 
\begin{align}
    {\cal F}^{(1),3,\tau(2)}_{ggg,\rm fin} \neq {\cal F}^{(1),3,\tau(2)}_{ggg,\rm fin,\rm QCD}\,.
\end{align}
The Casimirs in fundamental and adjoint representations are denoted as $C_F=(N^2-1)/(2N)$ and $C_A=N$, respectively, $T_F=1/2$ is the conventional normalisation factor and $n_f$ is the number of massless fermionic flavors in QCD. This indicates a clear violation of the PMT. This non-matching may be attributed to the incomplete factorisation of the leading order amplitude from the one loop form factor of EM tensor in QCD. 

In ${\cal N}=4$ SYM, it has been observed~\cite{Brandhuber:2012vm,Banerjee:2016kri}, that the coefficients of the MPLs in HT term at any order, both in the poles as well as the finite part become independent of the kinematics only after dividing the loop matrix elements by their leading order contribution. This independence of kinematics can also be seen for HT terms in QCD. For the poles, one can understand this factorisation of kinematics in terms of their universal nature, but it is not trivial to reason out why it happens for the finite part too. For example, the half-BPS operator in ${\cal N}=4$ SYM which  has been seen to contain only HT terms up to two loop order~\cite{Brandhuber:2012vm,Banerjee:2016kri} in the perturbation theory, matches with the HT contribution of the Konishi operator~\cite{Banerjee:2016kri} (with $g\phi\phi$ as external particles) because of the exact factorisation of the leading order contributions from both sides, in the poles as well as finite parts. 
For the current process (a) in QCD, the leading order does not factorise from the HT part of one- and two-loop, see Eq. (4.3) in~\cite{Ahmed:2014gla}, which is in sharp contrast to the counterpart process in ${\cal N}=4$ SYM. Consequently, the mismatch of the HT terms occur.

In addition, it is also observed that in the finite term of the FF if we take the double collinear limit $y \rightarrow 0$  and $z \rightarrow 0$ (which is equivalent to single soft limit), the residual kinematics factorises. Moreover, it turns out to be identical to the ${\cal N}=4$ SYM counterparts upon taking the same limit on latter part.



The same things get repeated for the process (b) i.e. violation of PMT. However, in contrast to the process (a), the HT terms for the finite part matches partially with the ${\cal N}=4$ SYM counterpart, while the rest of them do match only in the double collinear limit. More precisely, at one-loop the terms proportional to quantities like $\zeta_{2},\log(z) \log(y)$ and $\log(z)\log(1-z)$ match with the ${\cal N}=4$ SYM result. But for the remaining terms, we need to take the double collinear or single soft limit appropriately to get the agreement. 

In conclusion, for the first time the PMT is found to break down for three-point FFs arising from the EM tensor.

\section{Checks on results}
\label{sec:checks}
We perform a number of consistency checks in order to ensure the reliability of our results.
\begin{itemize}
    \item All the three-point FFs at one- and two-loop and two-point FFs up to three-loops exhibit the correct universal infrared behaviour depicted in section~\ref{ss:ir} which serves as the most stringent check on our results.
    

    \item Appearance of correct IR poles, though, is a powerful check, it does not fully ensure the correctness of the finite term. Often one needs to perform some additional checks on finite parts. For the processes under consideration, we check the kinematics symmetry in finite FF as well as remainder functions. For the processes with three gluons, the result should be symmetric under mutual exchange of momentum of any of the two gluons: $p_i \leftrightarrow p_j$ i.e. $x_i \leftrightarrow x_j$ where $x_1=x,x_2=z,x_3=y$. Similarly, $x_1 \leftrightarrow x_3$ symmetry must be there for the processes with final state particles containing $\phi(p_2)\phi(p_3)$ or $\lambda(p_2){\bar\lambda}(p_3)$. For all the processes, we have checked the kinematics symmetry in finite FF as well as remainders numerically with the help of \ginac~\cite{Bauer:2000cp,Vollinga:2004sn} through \polylogtools~\cite{Duhr:2019tlz}. Indeed, they exhibit the expected symmetry which in turn raises the reliability of the final results. 
    %
    %
    For all the other processes, this symmetry is checked and found to hold true except the ${\tau(0)}$ term of the ${\cal O}^1$ for the process with external particles $\phi\lambda\bar\lambda$ at two-loop: ${\cal F}^{(2),2,\tau(0)}_{\phi\lambda\bar\lambda,\text{fin}}$ and ${\mathcal R}^{(2),2,\tau(0)}_{\phi\lambda\bar\lambda}$. Owing to the presence of Levi-Civita tensor in the operator, we cannot always demand the fulfillness of the symmetry mentioned above.

    \item The anomalous dimensions for the mixing matrix in \eqref{eq:ZMat} between the operators ${\cal O}^1_B$ and ${\cal O}^1_F$ agree with the the partially available two-loop results~\cite{Brandhuber:2016fni}. Moreover, the anomalous dimensions of the operator ${\cal O}^2$ in \eqref{eq:oper-renorm-ITO-mixing} are identical to the Konishi supermultiplet, in accordance with the expectations~\cite{Eden:2005ve,Brandhuber:2016fni}. 
    
\end{itemize}

\section{Conclusions}
\label{sec:concl}
In this article, we have computed several two- and three-point form factors (FFs) to three- and two-loops, respectively, in maximally supersymmetric Yang Mills theory for three different choices of local gauge invariant operators by following Feynman diagrammatic approach. We have analysed several processes with different on-shell states.

We have considered two different operators~\cite{Intriligator:1999ff,Bianchi:2001cm,Beisert:2003ys,Eden:2005ve,Eden:2009hz,Brandhuber:2016fni} which are constructed out of non-protected dimension-three (classical)  fermionic and scalar components belonging to SU($2|3$) closed sub-sector of ${\cal N}=4$ SYM. The first one $({\cal O}^1)$ is a conformal descendant of the half-BPS operator, whereas the second one $({\cal O}^2)$ is a supersymmetric descendant of the Konishi primary. By exploiting the universal structures of the infrared divergences~\cite{Ahmed:2015qpa,Banerjee:2018yrn} associated to the FFs, the mixing among the non-protected fermionic and bosonic components of these operators are analysed up to three-loops in perturbation theory. Corresponding to the operator ${\cal O}^1$, the ultraviolet (UV)  anomalous dimensions or equivalently, the quantum corrections to the dilatation operators are computed fully to two-loops and partially at three-loop level. Similarly, corresponding to the operator $({\cal O}^2)$, we have computed partially the UV anomalous dimensions to two-loops. Through our computations, the anomalous dimensions of ${\cal O}^1$ are now fully available to two-loops and partially at three-loop. Our results of the anomalous dimensions of ${\cal O}^1$ are consistent with the partially available results to two-loops~\cite{Brandhuber:2016fni}. Moreover, the anomalous dimensions of ${\cal O}^2$ are identical to the Konishi supermultiplet, in accordance with the expectations~\cite{Eden:2005ve,Brandhuber:2016fni}.

We have analysed the results of the FFs from the perspective of principle of uniform and maximal transcendentality. The energy-momentum (EM) tensor behaves exactly like the half-BPS operator. These are uniform transcendental (UT), and the highest transcendental (HT) weight terms of the FFs and finite remainders are same as that of half-BPS operator. This is in accordance with our classical expectation: since the stress-tensor is protected and lies in the same multiplet as the half-BPS, an exact SUSY Ward identity would relate these two FFs. The most interesting finding of our analysis is that, the three-point FFs of the EM tensor violate the principle of maximal transcendentality (PMT). If we take the three-point quark and gluon FFs of EM tensor in QCD~\cite{Ahmed:2014gla,Ahmed:2016yox} and change the fermions from fundamental to adjoint representation, the HT weight terms do not match with the corresponding results in ${\cal N}=4$ SYM. This is a clear violation of the PMT and it is observed for the first time for three-point FF.

We find that except the three-point one-loop FF of operator ${\cal O}^1$, which is a conformal descendant of the half-BPS, with $\phi\phi\phi$ on-shell states, none of the other two- and three-point FFs of ${\cal O}^1$ exhibit UT behaviour. The general observation is that the FFs of SUSY protected operators exhibit UT behaviour. However, there is no clear indication whether the reverse is also true. Through the results depicted in \eqref{eq:FF-O1-ppp-UT}, we see that the FF of an unprotected operator, such as ${\cal O}^1$, can even be UT. The HT parts of the three-point remainder function and the finite FFs at one- as well as two-loop are independent of the external legs and they are identical to that of half-BPS.

None of the FFs and finite remainders of operator ${\cal O}^2$, which is a supersymmetric descendant of the Konishi primary, exhibits UT behaviour. In contrast to ${\cal O}^1$, the HT weight terms of the three-point FFs of ${\cal O}^2$ do depend on the nature of external on-shell states. Similar behaviour is observed for the three-point FFs of Konishi operator in ref.~\cite{Banerjee:2016kri}. So, we see that the dependence of HT terms on the nature of external on-shell states is a general feature of the FFs of Konishi primary and its descendants. Moreover, none of the HT terms of the FFs of ${\cal O}^2$ matches with that of half-BPS. 

In ref.~\cite{Brandhuber:2014ica}, the minimal FFs of long BPS operators (more than two fields) are computed to two-loops, and the corresponding HT piece is found to be a universal in all FFs with long (more than two-fields), unprotected operators. For the FF of ${\cal O}^2$ with the external on-shell states $\phi\phi\phi$, the one-loop contribution comes only from the scalar part ${\cal O}^2_B$, the fermionic part ${\cal O}^2_F$ does not contribute at this loop level. So, this one-loop FF effectively comes from a long, unprotected operator. For this case, we find that the HT part of the one-loop FF does not match with the HT term of long BPS operator, and consequently it violates the conjecture.

\section*{Acknowledgements}
We thank A. Brandhuber and G. Travaglini for their suggestions and comments on the manuscript. We also thank A. H. Ajjath, P. Mukherjee, S. Seth and PB thanks A. Signer for discussions.


\appendix
\section{Results}
\label{app:results}
In this appendix, we present the results of all the two- and three-point FFs, and finite remainders by setting $\mu^2=-q^2$. The $\mu$ dependence can be restored through renormalisation group equation. The results are presented in terms of multiple polylogarithms~\cite{Vollinga:2004sn}. We begin by presenting the results\footnote{In article~\cite{Banerjee:2016kri}, the results are presented in terms of two-dimensional harmonic polylogarithms which we have converted to multiple polylogarithms for convenience.} of the half-BPS operator~\cite{vanNeerven:1985ja,Gehrmann:2011xn,Brandhuber:2012vm,Banerjee:2016kri} because of its relevance in the present context.

\subsection{Two- and three-point form factors of half-BPS operator}
\label{app:results-hBPS}
%
\begin{align}
\begin{autobreak}
    {\cal F}^{(1),\text{half-BPS}}_{{\cal C}_1{\cal C}_2,\rm fin} =
\epsilon^4 \bigg( \frac{949  \zeta_{2}^3}{4480}
-\frac{49  \zeta_{3}^2}{144}
\bigg) 
+\epsilon^3 \bigg( \frac{7}{24}  \zeta_{2} \zeta_{3}
-\frac{31  \zeta_{5}}{20}
\bigg)
+ \epsilon^2 \bigg( \frac{47 \zeta_{2}^2}{80}
\bigg)
-\frac{8}{\epsilon^2}
+ \epsilon \bigg( -\frac{7  \zeta_{3}}{3}
\bigg)
+\zeta_{2}
\,,
\end{autobreak}\\
 \begin{autobreak}
\label{eq:hBPS-CatFin-1L}
    {\cal F}^{(1),\text{half-BPS}}_{{\cal C}_1{\cal C}_2{\cal C}_3,\rm fin} =
    -\frac{\pi ^2}{6}
    -G(0,y) G(0,z)
    -G(0,y) G(1,z)
    -G(0,z) G(1-z,y)
    +2 G(1,z) G(-z,y)
    +G(0,1,z)
    -G(0,1-z,y)
    +2 G(1,0,y)
    +G(1,0,z)
    -G(1-z,0,y)
    +2 G(-z,1-z,y)\,
\end{autobreak}\\
\begin{autobreak}
\label{eq:hBPS-CatFin-2L}
    {\cal F}^{(2),\text{half-BPS}}_{{\cal C}_1{\cal C}_2{\cal C}_3,\rm fin} =
\frac{49 \pi ^4}{720}
+2 G(0,0,y) G(0,0,z)
+2 G(0,0,y) G(0,1,z)
+2 G(0,0,z) G(0,1-z,y)
-2 G(0,1,z) G(0,-z,y)
+2 G(1,0,y) G(0,1,z)
+2 G(0,0,y) G(1,0,z)
-2 G(1,0,z) G(0,-z,y)
+2 G(1,0,y) G(1,0,z)
+2 G(0,0,y) G(1,1,z)
-4 G(1,1,z) G(0,-z,y)
+2 G(0,0,z) G(1-z,0,y)
-4 G(0,1,z) G(1-z,0,y)
-4 G(1,0,z) G(1-z,0,y)
+2 G(0,0,z) G(1-z,1-z,y)
-4 G(0,1,z) G(1-z,1-z,y)
+4 G(1,0,z) G(1-z,1-z,y)
+6 G(0,1,z) G(1-z,-z,y)
-2 G(1,0,z) G(1-z,-z,y)
+2 G(0,1,z) G(-z,0,y)
+2 G(1,0,z) G(-z,0,y)
-4 G(1,1,z) G(-z,0,y)
+\pi ^2 \bigg(\frac{1}{2} G(0,y) G(0,z)
+\frac{1}{2} G(0,z) G(1-z,y)
+\frac{1}{2} G(0,y) G(1,z)
-\frac{4}{3} G(1,z) G(1-z,y)
+\frac{1}{3} G(1,z) G(-z,y)
+\frac{1}{2} G(0,1-z,y)
+\frac{1}{2} G(1-z,0,y)
-\frac{4}{3} G(1-z,1-z,y)
+\frac{1}{3} G(-z,1-z,y)
-\frac{4}{3} G(0,1,y)
-G(1,0,y)
+\frac{4}{3} G(1,1,y)
-\frac{1}{2} G(0,1,z)
-\frac{1}{2} G(1,0,z)\bigg)
+4 G(0,1,z) G(-z,1-z,y)
-4 G(1,0,z) G(-z,1-z,y)
-8 G(0,1,z) G(-z,-z,y)
+8 G(1,1,z) G(-z,-z,y)
-2 G(0,y) G(0,0,1,z)
+6 G(0,0,1,z) G(1-z,y)
-8 G(0,0,1,z) G(-z,y)
-2 G(0,z) G(0,0,1-z,y)
+2 G(1,z) G(0,0,1-z,y)
+4 G(1,z) G(0,0,-z,y)
-2 G(0,1,0,y) G(0,z)
-2 G(0,1,0,y) G(1,z)
-2 G(0,y) G(0,1,0,z)
-2 G(0,1,0,z) G(1-z,y)
-2 G(0,y) G(0,1,1,z)
+4 G(0,1,1,z) G(-z,y)
-2 G(0,z) G(0,1-z,0,y)
+2 G(1,z) G(0,1-z,0,y)
+2 G(0,z) G(0,1-z,1-z,y)
-2 G(1,z) G(0,1-z,-z,y)
+4 G(1,z) G(0,-z,0,y)
-2 G(0,z) G(0,-z,1-z,y)
-4 G(1,z) G(0,-z,1-z,y)
-4 G(1,0,0,y) G(0,z)
-4 G(1,0,0,y) G(1,z)
-2 G(0,y) G(1,0,0,z)
-2 G(1,0,0,z) G(1-z,y)
-2 G(0,y) G(1,0,1,z)
+4 G(1,0,1,z) G(-z,y)
+2 G(0,z) G(1,0,1-z,y)
-2 G(0,y) G(1,1,0,z)
+8 G(1,1,0,z) G(1-z,y)
-4 G(1,1,0,z) G(-z,y)
+2 G(0,z) G(1,1-z,0,y)
+2 G(0,z) G(1-z,0,0,y)
+2 G(1,z) G(1-z,0,0,y)
-2 G(0,z) G(1-z,0,1-z,y)
-2 G(1,z) G(1-z,0,-z,y)
-2 G(0,z) G(1-z,1,0,y)
+4 G(1,z) G(1-z,1,0,y)
-2 G(0,z) G(1-z,1-z,0,y)
-4 G(1,z) G(1-z,1-z,-z,y)
-2 G(1,z) G(1-z,-z,0,y)
-2 G(0,z) G(1-z,-z,1-z,y)
+8 G(1,z) G(1-z,-z,-z,y)
+2 G(0,z) G(-z,0,1-z,y)
-4 G(1,z) G(-z,0,1-z,y)
+2 G(0,z) G(-z,1-z,0,y)
-4 G(1,z) G(-z,1-z,0,y)
-4 G(0,z) G(-z,1-z,1-z,y)
+8 G(1,z) G(-z,1-z,-z,y)
+8 G(1,z) G(-z,-z,1-z,y)
-8 G(1,z) G(-z,-z,-z,y)
-4 G(0,0,1,0,y)
+2 G(0,0,1,1,z)
+2 G(0,0,1-z,1-z,y)
+4 G(0,0,-z,1-z,y)
+2 G(0,1,0,1,z)
-2 G(0,1,0,1-z,y)
+8 G(0,1,1,0,y)
+2 G(0,1,1,0,z)
-2 G(0,1,1-z,0,y)
+2 G(0,1-z,0,1-z,y)
-2 G(0,1-z,1,0,y)
+2 G(0,1-z,1-z,0,y)
-2 G(0,1-z,-z,1-z,y)
+4 G(0,-z,0,1-z,y)
+4 G(0,-z,1-z,0,y)
-4 G(0,-z,1-z,1-z,y)
+2 G(1,0,0,1,z)
-4 G(1,0,0,1-z,y)
+8 G(1,0,1,0,y)
+2 G(1,0,1,0,z)
-4 G(1,0,1-z,0,y)
+8 G(1,1,0,0,y)
+2 G(1,1,0,0,z)
-8 G(1,1,1,0,y)
-4 G(1,1-z,0,0,y)
+2 G(1-z,0,0,1-z,y)
-2 G(1-z,0,1,0,y)
+2 G(1-z,0,1-z,0,y)
-2 G(1-z,0,-z,1-z,y)
-4 G(1-z,1,0,0,y)
+4 G(1-z,1,0,1-z,y)
+4 G(1-z,1,1-z,0,y)
+2 G(1-z,1-z,0,0,y)
+4 G(1-z,1-z,1,0,y)
-4 G(1-z,1-z,-z,1-z,y)
-2 G(1-z,-z,0,1-z,y)
-2 G(1-z,-z,1-z,0,y)
+8 G(1-z,-z,-z,1-z,y)
-4 G(-z,0,1-z,1-z,y)
-4 G(-z,1-z,0,1-z,y)
-4 G(-z,1-z,1-z,0,y)
+8 G(-z,1-z,-z,1-z,y)
+8 G(-z,-z,1-z,1-z,y)
-8 G(-z,-z,-z,1-z,y)
+\zeta_3 \Big(-G(1-z,y)-G(0,y)-G(0,z)-G(1,z)\Big)\,.
\end{autobreak}\\
\begin{autobreak}
\label{eq:hBPS-FinRem}
    \mathcal{R}^{(2),\text{half-BPS}}_{{\cal C}_1{\cal C}_2{\cal C}_3} =    
    -4 G(0,1,z) G(1-z,0,y)
-4 G(1,0,z) G(1-z,0,y)
-4 G(0,1,z) G(1-z,1-z,y)
+4 G(1,0,z) G(1-z,1-z,y)
+8 G(0,1,z) G(1-z,-z,y)
+4 G(0,1,z) G(-z,0,y)
+4 G(1,0,z) G(-z,0,y)
+4 G(0,1,z) G(-z,1-z,y)
-4 G(1,0,z) G(-z,1-z,y)
+\pi ^2 \bigg(-\frac{4}{3} G(1,z) G(1-z,y)
+\frac{4}{3} G(1,z) G(-z,y)
-\frac{4}{3} G(1-z,1-z,y)
+\frac{4}{3} G(-z,1-z,y)
-\frac{4}{3} G(0,1,y)
+\frac{4}{3} G(1,1,y)\bigg)
-8 G(0,1,z) G(-z,-z,y)
+8 G(0,0,1,z) G(1-z,y)
-8 G(0,0,1,z) G(-z,y)
-4 G(0,z) G(0,0,1-z,y)
+4 G(1,z) G(0,0,-z,y)
-4 G(0,z) G(0,1-z,0,y)
+4 G(1,z) G(0,-z,0,y)
+4 G(0,z) G(1,0,1-z,y)
-4 G(1,z) G(1,0,-z,y)
+8 G(1,1,0,z) G(1-z,y)
-8 G(1,1,0,z) G(-z,y)
+4 G(0,z) G(1,1-z,0,y)
-4 G(1,z) G(1,-z,0,y)
-4 G(0,z) G(1-z,0,1-z,y)
+4 G(1,z) G(1-z,1,0,y)
-4 G(0,z) G(1-z,1-z,0,y)
-4 G(1,z) G(1-z,1-z,-z,y)
+8 G(1,z) G(1-z,-z,-z,y)
+4 G(0,z) G(-z,0,1-z,y)
-4 G(1,z) G(-z,1,0,y)
+4 G(0,z) G(-z,1-z,0,y)
+4 G(1,z) G(-z,1-z,-z,y)
-8 G(1,z) G(-z,-z,-z,y)
-4 G(0,0,1,0,y)
+4 G(0,0,-z,1-z,y)
+8 G(0,1,1,0,y)
+4 G(0,-z,0,1-z,y)
+4 G(0,-z,1-z,0,y)
+4 G(1,0,1,0,y)
-4 G(1,0,-z,1-z,y)
-8 G(1,1,1,0,y)
-4 G(1,-z,0,1-z,y)
-4 G(1,-z,1-z,0,y)
+4 G(1-z,1,0,1-z,y)
+4 G(1-z,1,1-z,0,y)
+4 G(1-z,1-z,1,0,y)
-4 G(1-z,1-z,-z,1-z,y)
+8 G(1-z,-z,-z,1-z,y)
-4 G(-z,1,0,1-z,y)
-4 G(-z,1,1-z,0,y)
-4 G(-z,1-z,1,0,y)
+4 G(-z,1-z,-z,1-z,y)
-8 G(-z,-z,-z,1-z,y)\,.
\end{autobreak}
\end{align}

\subsection{\texorpdfstring{Two-point form factors of operator $\mathcal{O}^1$}{}}
\label{app:results-O1-2pt}
The results of the Sudakov FFs of operator ${\cal O}^1$ are presented below:
\begin{align}
\label{eq:res-O2-ll-1loop}
\begin{autobreak}
{\cal F}^{(1),1}_{\lambda\bar \lambda} =
\epsilon^4 \bigg( \frac{949  \zeta_{2}^3}{4480}
+\frac{47  \zeta_{2}^2}{320}
+\frac{ \zeta_{2}}{4}
-\frac{49  \zeta_{3}^2}{144}
+\frac{7  \zeta_{3}}{12}
-2 \bigg) 
+\epsilon^3 \bigg( \frac{7}{24}  \zeta_{2} \zeta_{3}
-\frac{ \zeta_{2}}{4}
-\frac{7  \zeta_{3}}{12}
-\frac{31  \zeta_{5}}{20}
+2  \bigg)
+ \epsilon^2 \bigg( \frac{47 \zeta_{2}^2}{80}
+\frac{\zeta_{2}}{4}
-2 \bigg)
-\frac{8}{\epsilon^2}
+ \epsilon \bigg( -\frac{7  \zeta_{3}}{3}
+2 \bigg)
+\zeta_{2}
-2
 \,,
\end{autobreak}\\
\begin{autobreak}
{\cal F}^{(2),1}_{\lambda\bar \lambda} =
\frac{32}{\epsilon^4}
+\epsilon^2 \bigg(\frac{2313 \zeta_{2}^3}{280}
+\frac{3 \zeta_{2}^2}{2}
-8 \zeta_{2}
+\frac{901 \zeta_{3}^2}{36}
+\frac{7 \zeta_{3}}{6}
+\frac{439}{4}\bigg)
+ \frac{1}{\epsilon^2}\Big( 16-4 \zeta_{2} \Big)
+\epsilon \bigg(-\frac{23 \zeta_{2} \zeta_{3}}{6}
+4 \zeta_{2}
-\frac{8 \zeta_{3}}{3}
-\frac{71 \zeta_{5}}{10}
-\frac{113}{2}\bigg)
+\frac{1}{\epsilon} \bigg(\frac{50 \zeta_3}{3}-16 \bigg)
-\frac{21 \zeta_{2}^2}{5}
-4 \zeta_{2}
+28
\,,\nonumber
\end{autobreak}\\
\begin{autobreak}
{\cal F}^{(3),1}_{\lambda\bar \lambda} =
-\frac{256}{3
    \epsilon^6}
-\frac{64}{\epsilon^4}
+\frac{1}{\epsilon^3} \bigg(64-\frac{176 \zeta_{3}}{3}\bigg)
+\frac{1}{\epsilon^2} \bigg( \frac{494 \zeta_{2}^2}{45} +16 \zeta_{2}-160 \bigg)
    +\frac{1}{\epsilon} \bigg( \frac{170 \zeta_{2} \zeta_{3}}{9}
    -16 \zeta_{2}
    +44
    \zeta_{3}
    +\frac{1756 \zeta_{5}}{15}
    +388 \bigg)
    -\frac{22523 \zeta_{2}^3}{270}
    -\frac{167 \zeta_{2}^2}{10}
    +60 \zeta_{2}
    -\frac{1766 \zeta_{3}^2}{9}
    -44
    \zeta_{3}-710
\,.
\end{autobreak}
\end{align}


\subsection{\texorpdfstring{Three-point form factors of operator ${\cal O}^{1}$}{}}
\label{app:results-O1-3pt}
The HT parts of one- as well as two-loop results for the $\mathcal{O}^1$ operator are same as that of half-BPS:
\begin{align}
        \label{eq:res-hBPS2-HT}
        &{\cal F}^{(n),1,\tau(2n)}_{\lambda\bar\lambda,\rm fin} = 
        {\cal F}^{(n),\text{half-BPS}}_{{\cal C}_1{\cal C}_2,\rm fin}\,,\nonumber\\
        &{\cal F}^{(n),1,\tau(2n)}_{g\lambda\bar\lambda,\rm fin}  =
        {\cal F}^{(n),1,\tau(2n)}_{\phi\lambda\bar\lambda,\rm fin} = 
        {\cal F}^{(n),1,\tau(2n)}_{\phi\phi \phi,\rm fin} 
        \equiv{\cal F}^{(n),1, \tau(2n)}_{{\cal C}_1{\cal C}_2{\cal C}_3,\rm fin} = 
        {\cal F}^{(n),\text{half-BPS}}_{{\cal C}_1{\cal C}_2{\cal C}_3,\rm fin}\,,\nonumber\\
        &{\mathcal R}^{(n),1,\tau(2n)}_{g\lambda\bar\lambda} = 
        {\mathcal R}^{(n),1,\tau(2n)}_{\phi\lambda\bar\lambda}=
         {\mathcal R}^{(n),1,\tau(2n)}_{\phi\phi \phi}
        \equiv{\mathcal R}^{(n),1, \tau(2n)}_{{\cal C}_1{\cal C}_2{\cal C}_3} = 
        {\mathcal R}^{(n),\text{half-BPS}}_{{\cal C}_1{\cal C}_2{\cal C}_3}\,.
\end{align}
The lower transcendental weight terms at one-loop are same irrespective of the external states
\begin{align}
        \label{eq:res-hBPS2-1Loop-lowerT}
        &{\cal F}^{(1),1,\tau(1)}_{g\lambda\bar\lambda,\rm fin} =
        {\cal F}^{(1),1,\tau(1)}_{\phi\lambda\bar\lambda,\rm fin} = 
        {\cal F}^{(1),1,\tau(1)}_{\phi\phi\phi,\rm fin} = 0\,,\\ \nonumber\\
        &{\cal F}^{(1),1,\tau(0)}_{g\lambda\bar\lambda,\rm fin} =
        {\cal F}^{(1),1,\tau(0)}_{\phi\lambda\bar\lambda,\rm fin} = -2\,\nonumber\\
        &{\cal F}^{(1),1,\tau(0)}_{\phi\phi\phi,\rm fin} = 0\,.
\end{align}
The lower weight terms at two-loop are obtained as
\begin{align}
\begin{autobreak}
{\cal F}^{(2),1,\tau(3)}_{g\lambda\bar\lambda,\rm fin} =
{\cal F}^{(2),1,\tau(3)}_{\phi\lambda\bar\lambda,\rm fin} =
{\cal F}^{(2),1,\tau(3)}_{\phi\phi\phi,\rm fin} = 
0\,,
\end{autobreak}\\ \nonumber\\
%
\begin{autobreak}
{\cal F}^{(2),1,\tau(2)}_{g\lambda\bar\lambda,\rm fin} ={\cal F}^{(2),1,\tau(2)}_{\phi\lambda\bar\lambda,\rm fin} =
2 G(0,y) (G(0,z)
+G(1,z))
+2 G(0,z) G(1-z,y)
-4 G(1,z) G(-z,y)
+2 G(0,1-z,y)
+2 \
G(1-z,0,y)
-4 G(-z,1-z,y)
-4 G(1,0,y)
-2 G(0,1,z)
-2 G(1,0,z)
+\frac{\pi ^2}{3}\,,\nonumber
\end{autobreak}\\
\begin{autobreak}
{\cal F}^{(2),1,\tau(2)}_{\phi\phi\phi,\rm fin} =
2 G(0,y) G(0,z)
+2 G(0,y) G(1,z)
+2 G(0,y) G(1-z,y)
+2 G(0,z) G(1-z,y)
-4 G(1,z) G(-z,y)
-4 G(-z,1-z,y)
-4 G(1,0,y)
-2 G(0,z) G(1,z)
+2 \zeta_{2}\,,
\end{autobreak}\\ \nonumber \\
%
\begin{autobreak}
{\cal F}^{(2),1,\tau(1)}_{g\lambda\bar\lambda,\rm fin} =
{\cal F}^{(2),1,\tau(1)}_{\phi\lambda\bar\lambda,\rm fin} =
0\,,\nonumber
\end{autobreak}\\
\begin{autobreak}
{\cal F}^{(2),1,\tau(1)}_{\phi\phi\phi,\rm fin} =
4 G(1-z,y)+4 G(0,y)+4 G(0,z)+4 G(1,z)\,,
\end{autobreak}\\ \nonumber \\
%
\begin{autobreak}
{\cal F}^{(2),1,\tau(0)}_{\phi\lambda\bar\lambda,\rm fin} =
1
-\frac{4 y}{2 y+z-1}\,,\nonumber
\end{autobreak}\\
\begin{autobreak}
{\cal F}^{(2),1,\tau(0)}_{g\lambda\bar\lambda,\rm fin} =
12
-\frac{4 y}{2 y+z-1}\,,\nonumber
\end{autobreak}\\
\begin{autobreak}
{\cal F}^{(2),1,\tau(0)}_{\phi\phi\phi,\rm fin} =
-30\,,
\end{autobreak} \\ \nonumber \\
%
\begin{autobreak}
{\mathcal R}^{(2),{1},\tau(3)}_{g\lambda\bar\lambda} =
{\mathcal R}^{(2),{1},\tau(3)}_{\phi\lambda\bar\lambda} =
{\mathcal R}^{(2),{1},\tau(3)}_{\phi\phi\phi} =
0\,,
\end{autobreak}\\ \nonumber\\
%
\begin{autobreak}
{\mathcal R}^{(2),{1},\tau(2)}_{g\lambda\bar\lambda} =
{\mathcal R}^{(2),{1},\tau(2)}_{\phi\lambda\bar\lambda} =
-4 \zeta_2\,,\nonumber
\end{autobreak}\\
\begin{autobreak}
{\mathcal R}^{(2),{1},\tau(2)}_{\phi\phi\phi} =
2 G(0,y) G(0,z)
+2 G(0,y) G(1,z)
+2 G(0,y) G(1-z,y)
+2 G(0,z) G(1-z,y)
-4 G(1,z) G(-z,y)
-4 G(-z,1-z,y)
-4 G(1,0,y)
-2 G(0,z) G(1,z)
+2 \zeta_{2}\,,
\end{autobreak}\\ \nonumber \\
%
\begin{autobreak}
{\mathcal R}^{(2),{1},\tau(1)}_{g\lambda\bar\lambda} =
{\mathcal R}^{(2),{1},\tau(1)}_{\phi\lambda\bar\lambda} =
0\,,\nonumber
\end{autobreak}\\
\begin{autobreak}
{\mathcal R}^{(2),{1},\tau(1)}_{\phi\phi\phi} =
4 G(1-z,y)
+4 G(0,y)
+4 G(0,z)
+4 G(1,z)
\,,
\end{autobreak}\\ \nonumber \\
%
\begin{autobreak}
{\mathcal R}^{(2),{1},\tau(0)}_{g\lambda\bar\lambda} =
-1\,,\nonumber
\end{autobreak}\\
\begin{autobreak}
{\mathcal R}^{(2),{1},\tau(0)}_{\phi\lambda\bar\lambda} =
10
-\frac{4 y}{2 y+z-1}\,,
\end{autobreak}\nonumber\\
\begin{autobreak}
{\mathcal R}^{(2),{1},\tau(0)}_{\phi\phi\phi} =
-30\,.
\end{autobreak}\nonumber\\
\end{align}
%


\subsection{\texorpdfstring{Three-point form factors of operator ${\cal O}^{2}$}{}}
\label{app:results-O2-3pt}
None of the HT parts of one- and two-loop results of the operator $\mathcal{O}^2$ coincides with that of the half-BPS:
%
\begin{align}
                &{\cal F}^{(n),2,\tau(2n)}_{{\cal C}_1{\cal C}_2{\cal C}_3,\rm fin} \neq
        {\cal F}^{(n),\text{half-BPS}}_{{\cal C}_1{\cal C}_2{\cal C}_3,\rm fin}\,,\nonumber\\
        &{\mathcal R}^{(n),2,\tau(2n)}_{\phi\phi\phi} \neq
        {\cal R}^{(n),\text{half-BPS}}_{{\cal C}_1{\cal C}_2{\cal C}_3,\rm fin} \,.
\end{align}
They also depend on the nature of external on-shell states. The finite remainder at two-loop can not be defined properly for remaining processes due to the involvement of one-loop diagrams arising from ${\cal O}^2_F$. Note that the leading order contributions to the processes with on-shell particles $g\lambda\bar\lambda$, $\phi\lambda\bar\lambda$ are loop induced which arises from ${\cal O}^2_B$. The explicit results of the FFs at one-loop (leading order) are
%


\begin{align}
        \label{eq:res-O2-1Loop}
        &{\cal F}^{(1),2,\tau(2)}_{g\lambda\bar\lambda,\rm fin}= 
        {\cal F}^{(1),2,\tau(2)}_{\phi\lambda\bar\lambda,\rm fin}=
        {\cal F}^{(1),1,\tau(2)}_{\phi\phi\phi,\rm fin} =
        0\,,\\ \nonumber\\
        &{\cal F}^{(1),2,\tau(1)}_{g\lambda\bar\lambda,\rm fin}=
        {\cal F}^{(1),2,\tau(1)}_{\phi\lambda\bar\lambda,\rm fin} =
        0\,, \nonumber\\
        &{\cal F}^{(1),1,\tau(1)}_{\phi\phi\phi,\rm fin} =
         2 G(1-z,y)+2 G(0,y)+2 G(0,z)+2 G(1,z)
         \,,\\ \nonumber\\
        &{\cal F}^{(1),2,\tau(0)}_{g\lambda\bar\lambda,\rm fin} =-\frac{8 {\epsilon} (z-1)^2}{y (y+z-1)}-\frac{16 \left(z^2+1\right)}{y (y+z-1)}\,,\nonumber\\
        &{\cal F}^{(1),1,\tau(0)}_{\phi\lambda\bar\lambda,\rm fin} =
        \frac{ (2 y+z-1)^2}{y (y+z-1)} \big(4 {\epsilon}-24 \big) \,,\nonumber\\
        &{\cal F}^{(1),1,\tau(0)}_{\phi\phi\phi,\rm fin} = -12\,.
\end{align}
The results of the FFs at two-loop (next-to-leading order) are
\begin{align}
\begin{autobreak}
{\cal F}^{(2),2,\tau(4)}_{g\lambda\bar\lambda,\rm fin} =0
\,,\nonumber
\end{autobreak}\\
\begin{autobreak}
{\cal F}^{(2),2,\tau(4)}_{\phi\lambda\bar\lambda,\rm fin} =0
\,,\nonumber
\end{autobreak}\\
\begin{autobreak}
{\cal F}^{(2),2,\tau(4)}_{\phi\phi\phi,\rm fin} =
-2 \zeta_{2} G(0,y) G(0,z)
+\zeta_{2} G(0,y) G(1,z)
+\zeta_{2} G(0,z) G(1-z,y)
-2 \zeta_{2} G(1,z) G(1-z,y)
+\zeta_{2} G(0,1-z,y)
+\zeta_{2} G(1-z,0,y)
-2 \zeta_{2} G(1-z,1-z,y)
-3 \zeta_{3} G(1-z,y)
-G(0,0,y) G(0,0,z)
+G(0,0,y) G(0,1,z)
+G(0,0,z) G(0,1-z,y)
+G(0,1,z) G(0,1-z,y)
+G(0,0,y) G(1,0,z)
-2 G(1,0,z) G(0,1-z,y)
-G(0,0,y) G(1,1,z)
+G(0,0,z) G(1-z,0,y)
+G(0,1,z) G(1-z,0,y)
-2 G(1,0,z) G(1-z,0,y)
-G(0,0,z) G(1-z,1-z,y)
+3 G(1,0,z) G(1-z,1-z,y)
-2 G(0,y) G(0,0,1,z)
-G(0,0,1,z) G(1-z,y)
-2 G(0,z) G(0,0,1-z,y)
-G(1,z) G(0,0,1-z,y)
+3 G(1,z) G(0,0,-z,y)
+G(0,y) G(0,1,0,z)
-G(0,1,0,z) G(1-z,y)
+G(0,y) G(0,1,1,z)
+G(0,z) G(0,1-z,0,y)
-G(1,z) G(0,1-z,0,y)
+G(0,z) G(0,1-z,1-z,y)
+G(0,y) G(1,0,0,z)
-G(1,0,0,z) G(1-z,y)
+G(0,y) G(1,0,1,z)
-2 G(0,y) G(1,1,0,z)
+3 G(1,1,0,z) G(1-z,y)
+G(0,z) G(1-z,0,0,y)
-G(1,z) G(1-z,0,0,y)
+G(0,z) G(1-z,0,1-z,y)
-2 G(0,z) G(1-z,1-z,0,y)
-G(0,0,1-z,1-z,y)
+3 G(0,0,-z,1-z,y)
-G(0,1-z,0,1-z,y)
-G(0,1-z,1-z,0,y)
-G(1-z,0,0,1-z,y)
-G(1-z,0,1-z,0,y)
-G(1-z,1-z,0,0,y)
+3 G(1-z,1-z,1,0,y)
+\zeta_{2} G(0,0,y)
+\zeta_{2} G(0,0,z)
+\zeta_{2} G(0,1,z)
+\zeta_{2} G(1,0,z)
-2 \zeta_{2} G(1,1,z)
-3 \zeta_{3} G(1,z)
+3 G(0,0,0,1,z)
-G(0,0,1,1,z)
-G(0,1,0,1,z)
-G(0,1,1,0,z)
-G(1,0,0,1,z)
-G(1,0,1,0,z)
-G(1,1,0,0,z)
+3 G(1,1,1,0,z)
-\frac{57 \zeta_{2}^{2}}{20} \,,
\end{autobreak}\\ \nonumber \\
%
\begin{autobreak}
{\cal F}^{(2),2,\tau(3)}_{g\lambda\bar\lambda,\rm fin} =0
\,,\nonumber
\end{autobreak}\\
\begin{autobreak}
{\cal F}^{(2),2,\tau(3)}_{\phi\lambda\bar\lambda,\rm fin} =
\frac{1}{2 y+z-1} \bigg( {4 y \zeta_{2} G(0,y)}
+{4 y \zeta_{2} G(0,z)}
-{8 y \zeta_{2} G(1,z)}
+{4 z \zeta_{2} G(1-z,y)}
-{4 \zeta_{2} G(1-z,y)}
+{4 y G(0,y) G(0,0,z)}
+{4 y G(0,0,z) G(1-z,y)}
+{4 z G(0,0,z) G(1-z,y)}
-{4 G(0,0,z) G(1-z,y)}
-{4 y G(0,z) G(0,1-z,y)}
-{4 z
    G(0,z) G(0,1-z,y)}
+{4 G(0,z) G(0,1-z,y)}
+{4 y G(1,z) G(0,-z,y)}
+{4 z G(1,z) G(0,-z,y)}
-{4 G(1,z) G(0,-z,y)}
-{4 y G(0,y) G(1,0,z)}
-{4 z G(1,0,z) G(1-z,y)}
+{4 G(1,0,z) G(1-z,y)}
-{4 y G(0,z) G(1-z,0,y)}
-{4 y G(0,1,0,z)}
+{4 y G(0,-z,1-z,y)}
+{4 z
    G(0,-z,1-z,y)}
-{4 G(0,-z,1-z,y)}
-{4 y G(1,0,0,z)}
+{8 y G(1,1,0,z)}
+{4 y G(1-z,1,0,y)}
-{4 y \zeta_{3}} \bigg)
\,,
\end{autobreak}\\ 
\begin{autobreak}
{\cal F}^{(2),2,\tau(3)}_{\phi\phi\phi,\rm fin} =
-2 G(0,0,y) G(0,z)
-2 G(0,0,y) G(1,z)
-2 G(0,y) G(0,0,z)
-2 G(0,0,z) G(1-z,y)
+4 G(0,y) G(0,1,z)
+2 G(0,1,z) G(1-z,y)
-4 G(0,1,z) G(-z,y)
+4 G(0,z) G(0,1-z,y)
-2 G(1,z) G(0,1-z,y)
+4 G(0,y) G(1,0,z)
-2 G(1,0,z) G(1-z,y)
-2 G(0,y) G(1,1,z)
+4 G(1,1,z) G(-z,y)
+4 G(0,z) G(1-z,0,y)
-2 G(1,z) G(1-z,0,y)
-2 G(0,z) G(1-z,1-z,y)
+4 G(1,z) G(1-z,-z,y)
+4 G(1,z) G(-z,1-z,y)
-4 G(1,z) G(-z,-z,y)
-2 G(0,0,1-z,y)
-2 G(0,1-z,0,y)
-2 G(0,1-z,1-z,y)
-2 G(1-z,0,0,y)
-2 G(1-z,0,1-z,y)
-2 G(1-z,1-z,0,y)
+4 G(1-z,-z,1-z,y)
+4 G(-z,1-z,1-z,y)
-4 G(-z,-z,1-z,y)
-4 \zeta_{2} G(0,y)
+4 \zeta_{2} G(1,y)
+4 G(0,1,0,y)
+4 G(1,0,0,y)
-4 G(1,1,0,y)
-4 \zeta_{2} G(0,z)
+4 \zeta_{2} G(1,z)
-6 G(0,0,1,z)
+2 G(0,1,0,z)
+2 G(0,1,1,z)
+2 G(1,0,0,z)
+2 G(1,0,1,z)
-6 G(1,1,0,z)+
4 \zeta_{3} \,,
\end{autobreak}\\ \nonumber\\
%
\begin{autobreak}
{\cal F}^{(2),2,\tau(2)}_{g\lambda\bar\lambda,\rm fin} =
\frac{1}{y z^4+y^2 z^3-y z^3+y z^2+y^2 z-y z} \Big( {6 G(0,z) G(1-z,y) z^5}
-{6 G(1,z) G(-z,y) z^5}
-{6 G(-z,1-z,y) z^5}
-{y \zeta_{2} z^4}
-{y G(0,y) G(0,z) z^4}
-{y G(0,y) G(1,z) z^4}
+{y
    G(0,z) G(1,z) z^4}
-{y G(0,y) G(1-z,y) z^4}
+{11 y G(0,z) G(1-z,y) z^4}
-{6 G(0,z)
    G(1-z,y) z^4}
-{10 y G(1,z) G(-z,y) z^4}
+{6 G(1,z) G(-z,y) z^4}
+{2 y G(1,0,y)
    z^4}
-{10 y G(-z,1-z,y) z^4}
+{6 G(-z,1-z,y) z^4}
+{5 y^2 \zeta_{2} z^3}
+{y \zeta_{2} z^3}
+{5 y^2 G(0,y) G(0,z) z^3}
+{y G(0,y) G(0,z) z^3}
-{7 y^2 G(0,y) G(1,z) z^3}
+{y G(0,y) G(1,z) z^3}
-{5 y^2 G(0,z) G(1,z) z^3}
-{y G(0,z) G(1,z) z^3}
-{7 y^2 G(0,y) G(1-z,y) z^3}
+{y G(0,y) G(1-z,y) z^3}
+{5 y^2 G(0,z) G(1-z,y) z^3}
+{y G(0,z) G(1-z,y) z^3}
-{6 G(0,z) G(1-z,y) z^3}
+{2 y^2 G(1,z) G(-z,y) z^3}
-{2 y G(1,z) G(-z,y) z^3}
+{6 G(1,z) G(-z,y) z^3}
+{12 y^2 G(0,1,z) z^3}
+{2 y^2 G(1,0,y) z^3}
-{2 y G(1,0,y) z^3}
+{2 y^2 G(-z,1-z,y) z^3}
-{2 y G(-z,1-z,y) z^3}
+{6 G(-z,1-z,y) z^3}
+{6 y^2
    \zeta_{2} z^2}
-{y \zeta_{2} z^2}
+{6 y^2 G(0,y) G(0,z) z^2}
-{y G(0,y) G(0,z)
    z^2}
-{12 y^3 G(0,y) G(1,z) z^2}
+{18 y^2 G(0,y) G(1,z) z^2}
-{y G(0,y) G(1,z)
    z^2}
-{6 y^2 G(0,z) G(1,z) z^2}
+{y G(0,z) G(1,z) z^2}
-{12 y^3 G(0,y) G(1-z,y)
    z^2}
+{18 y^2 G(0,y) G(1-z,y) z^2}
-{y G(0,y) G(1-z,y) z^2}
+{6 y^2 G(0,z) G(1-z,y)
    z^2}
-{13 y G(0,z) G(1-z,y) z^2}
+{6 G(0,z) G(1-z,y) z^2}
+{12 y^3 G(1,z) G(-z,y)
    z^2}
-{24 y^2 G(1,z) G(-z,y) z^2}
+{14 y G(1,z) G(-z,y) z^2}
-{6 G(1,z) G(-z,y)
    z^2}
+{12 y^3 G(0,1,z) z^2}
-{12 y^2 G(0,1,z) z^2}
+{12 y^3 G(1,0,y) z^2}
-{24 y^2 G(1,0,y) z^2}
+{2 y G(1,0,y) z^2}
+{12 y^3 G(-z,1-z,y) z^2}
-{24 y^2 G(-z,1-z,y) z^2}
+{14 y G(-z,1-z,y) z^2}
-{6 G(-z,1-z,y) z^2}
-{y^2 \zeta_{2} z}
+{y \zeta_{2} z}
-{y^2 G(0,y) G(0,z) z}
+{y G(0,y)
    G(0,z) z}
-{6 y^4 G(0,y) G(1,z) z}
+{24 y^3 G(0,y) G(1,z) z}
-{19 y^2 G(0,y) G(1,z)
    z}
+{y G(0,y) G(1,z) z}
+{y^2 G(0,z) G(1,z) z}
-{y G(0,z) G(1,z) z}
-{6 y^4 G(0,y) G(1-z,y) z}
+{24 y^3 G(0,y) G(1-z,y) z}
-{19 y^2 G(0,y) G(1-z,y) z}
+{y G(0,y) G(1-z,y) z}
-{y^2 G(0,z) G(1-z,y) z}
+{y G(0,z) G(1-z,y) z}
+{6 y^4 G(1,z) G(-z,y) z}
-{24 y^3 G(1,z) G(-z,y) z}
+{20 y^2 G(1,z) G(-z,y) z}
-{2 y G(1,z) G(-z,y) z}
+{6 y^4 G(0,1,z) z}
-{24 y^3 G(0,1,z) z}
+{18 y^2
    G(0,1,z) z}
+{6 y^4 G(1,0,y) z}
-{24 y^3 G(1,0,y) z}
+{20 y^2 G(1,0,y) z}
-{2 y G(1,0,y) z}
+{6 y^4 G(-z,1-z,y) z}
-{24 y^3 G(-z,1-z,y) z}
+{20 y^2 G(-z,1-z,y) z}
-{2 y G(-z,1-z,y) z}
+{6 y^4 G(0,y) G(1,z)}
-{12
    y^3 G(0,y) G(1,z)}
+{6 y^2 G(0,y) G(1,z)}
+{6 y^4 G(0,y) G(1-z,y)}
-{12 y^3 G(0,y)
    G(1-z,y)}
+{6 y^2 G(0,y) G(1-z,y)}
-{6 y^4 G(1,z) G(-z,y)}
+{12 y^3 G(1,z)
    G(-z,y)}
-{6 y^2 G(1,z) G(-z,y)}
-{6 y^4 G(0,1,z)}
+{12 y^3 G(0,1,z)}{y z^4+y^2
    z^3-y z^3+y z^2+y^2 z-y z}
-{6 y^2 G(0,1,z)}
-{6 y^4 G(1,0,y)}
+{12 y^3 G(1,0,y)}
-{6 y^2 G(1,0,y)}
-{6 y^4 G(-z,1-z,y)}
+{12 y^3 G(-z,1-z,y)}
-{6 y^2
    G(-z,1-z,y)} \Big)
\,,\nonumber
\end{autobreak}\\
\begin{autobreak}
{\cal F}^{(2),2,\tau(2)}_{\phi\lambda\bar\lambda,\rm fin} =
\frac{1}{2 y+z-1} \Big( 10 y G(0,y) G(0,z)
-z G(0,y) G(0,z)
+G(0,y) G(0,z)
-2 y G(0,y) G(1,z)
-z G(0,y) G(1,z)
+G(0,y) G(1,z)
-10 y G(0,z) G(1,z)
+z G(0,z) G(1,z)
-G(0,z) G(1,z)
-2 y G(0,y) G(1-z,y)
-z G(0,y) G(1-z,y)
+G(0,y) G(1-z,y)
+10 y G(0,z) G(1-z,y)
+11 z G(0,z) G(1-z,y)
-11 G(0,z) G(1-z,y)
-8 y G(1,z) G(-z,y)
-10 z G(1,z) G(-z,y)
+10 G(1,z)
    G(-z,y)
+12 y G(0,1,z)
-8 y G(1,0,y)
+2 z G(1,0,y)
-2 G(1,0,y)
-8 y G(-z,1-z,y)
-10 z G(-z,1-z,y)
+10
    G(-z,1-z,y)
+10 y \zeta_{2}
-z \zeta_{2}
+\zeta_{2} \Big)\,,
\end{autobreak}\\ 
\begin{autobreak}
{\cal F}^{(2),2,\tau(2)}_{\phi\phi\phi,\rm fin} =
G(0,y) G(0,z)
+G(0,y) G(1,z)
+G(0,y) G(1-z,y)
+G(0,z) G(1-z,y)
+10 G(1,z) G(-z,y)
+10 G(-z,1-z,y)
+10 G(1,0,y)
+11 G(0,z) G(1,z)
-23 \zeta_{2}
\,,
\end{autobreak}\\ \nonumber\\
%
\begin{autobreak}
{\cal F}^{(2),2,\tau(1)}_{g\lambda\bar\lambda,\rm fin} =
\frac{1}{y z^3-z^3+y^2 z^2-y z^2+y z-z+y^2-y} \Big( {6 z G(0,y) y^3}
-{6 G(0,y) y^3}
-{6 z G(1,z) y^3}
+{6 G(1,z)
    y^3}
-{6 z G(1-z,y) y^3}
+{6 G(1-z,y) y^3}
+{18 z^2 G(0,y) y^2}
-{12 z G(0,y) y^2}
+{6 G(0,y) y^2}
-{6 z^2 G(0,z) y^2}
-{6 z
    G(0,z) y^2}
+{24 z G(1,z) y^2}
-{12 G(1,z) y^2}
+{24 z G(1-z,y) y^2}
-{12 G(1-z,y) y^2}
+{12 z^3 G(0,y) y}
-{6 z^2 G(0,y) y}
+{6 z G(0,y) y}
-{6 z^3 G(0,z) y}
+{6 z G(0,z) y}
+{6 z^3 G(1,z) y}
-{24 z G(1,z) y}
+{6 G(1,z) y}
+{6 z^3 G(1-z,y) y}
-{24 z G(1-z,y) y}
+{6 G(1-z,y) y}
+{6 z^3 G(0,z)}
+{6 z^2 G(0,z)}
-{6 z^3 G(1,z)}
+{6 z G(1,z)}
-{6 z^3 G(1-z,y)}
+{6 z G(1-z,y)} \Big)
\,,\nonumber
\end{autobreak}\\
\begin{autobreak}
{\cal F}^{(2),2,\tau(1)}_{\phi\lambda\bar\lambda,\rm fin} =
\frac{1}{2 y^3+3 z y^2-3 y^2+z^2 y-4 z y+y-z^2+z} \bigg(
{12 G(0,y) y^3}
+{12 G(1,z) y^3}
+{12 G(1-z,y) y^3}
+{24 i \pi  y^3}
+{24 z G(0,y) y^2}
-{12 G(0,y) y^2}
+{12 z G(1,z) y^2}
-{24 G(1,z) y^2}
+{12 z G(1-z,y) y^2}
-{24 G(1-z,y) y^2}
+{36 i \pi  z y^2}
-{36 i \pi  y^2}
+{12 z^2 G(0,y) y}
-{12 z G(0,y) y}
-{12 z G(1,z) y}
+{12 G(1,z) y}
-{12 z G(1-z,y) y}
+{12 G(1-z,y) y}
+{12 i \pi  z^2 y}
-{48 i \pi  z y}
+{12 i \pi  y}
-{12 i \pi  z^2}
+{12 i \pi  z}
\bigg)
\,,
\end{autobreak}\\ 
\begin{autobreak}
{\cal F}^{(2),2,\tau(1)}_{\phi\phi\phi,\rm fin} =
-34 G(1-z,y)-34 G(0,y)-34 G(0,z)-34 G(1,z)
\,,
\end{autobreak}\\ \nonumber\\
%
\begin{autobreak}
{\cal F}^{(2),2,\tau(0)}_{g\lambda\bar\lambda,\rm fin} =
-44\,,\nonumber
\end{autobreak}\\
\begin{autobreak}
{\cal F}^{(2),2,\tau(0)}_{\phi\lambda\bar\lambda,\rm fin} =
\frac{-106 y-55 z+55}{2 y+z-1}\,,
\end{autobreak}\\ 
\begin{autobreak}
{\cal F}^{(2),2,\tau(0)}_{\phi\phi\phi,\rm fin} =
186
\,,
\end{autobreak}\\ \nonumber \\
\begin{autobreak}
{\mathcal R}^{(2),{2},\tau(4)}_{\phi\phi\phi} =
-2 \zeta_{2} G(0,y) G(0,z)
+\zeta_{2} G(0,y) G(1,z)
+\zeta_{2} G(0,z) G(1-z,y)
-2 \zeta_{2} G(1,z) G(1-z,y)
+\zeta_{2} G(0,1-z,y)
+\zeta_{2} G(1-z,0,y)
-2 \zeta_{2} G(1-z,1-z,y)
-2 \zeta_{3} G(1-z,y)
-G(0,0,y) G(0,0,z)
+G(0,0,y) G(0,1,z)
+G(0,0,z) G(0,1-z,y)
+G(0,1,z) G(0,1-z,y)
+G(0,0,y) G(1,0,z)
-2 G(1,0,z) G(0,1-z,y)
-G(0,0,y) G(1,1,z)
+G(0,0,z) G(1-z,0,y)
+G(0,1,z) G(1-z,0,y)
-2 G(1,0,z) G(1-z,0,y)
-G(0,0,z) G(1-z,1-z,y)
+3 G(1,0,z) G(1-z,1-z,y)
-2 G(0,y) G(0,0,1,z)
-G(0,0,1,z) G(1-z,y)
-2 G(0,z) G(0,0,1-z,y)
-G(1,z) G(0,0,1-z,y)
+3 G(1,z) G(0,0,-z,y)
+G(0,y) G(0,1,0,z)
-G(0,1,0,z) G(1-z,y)
+G(0,y) G(0,1,1,z)
+G(0,z) G(0,1-z,0,y)
-G(1,z) G(0,1-z,0,y)
+G(0,z) G(0,1-z,1-z,y)
+G(0,y) G(1,0,0,z)
-G(1,0,0,z) G(1-z,y)
+G(0,y) G(1,0,1,z)
-2 G(0,y) G(1,1,0,z)
+3 G(1,1,0,z) G(1-z,y)
+G(0,z) G(1-z,0,0,y)
-G(1,z) G(1-z,0,0,y)
+G(0,z) G(1-z,0,1-z,y)
-2 G(0,z) G(1-z,1-z,0,y)
-G(0,0,1-z,1-z,y)
+3 G(0,0,-z,1-z,y)
-G(0,1-z,0,1-z,y)
-G(0,1-z,1-z,0,y)
-G(1-z,0,0,1-z,y)
-G(1-z,0,1-z,0,y)
-G(1-z,1-z,0,0,y)
+3 G(1-z,1-z,1,0,y)
+\zeta_{2} G(0,0,y)
+\zeta_{3} G(0,y)
+\zeta_{2} G(0,0,z)
+\zeta_{2} G(0,1,z)
+\zeta_{2} G(1,0,z)
-2 \zeta_{2} G(1,1,z)
+\zeta_{3} G(0,z)
-2 \zeta_{3} G(1,z)
+3 G(0,0,0,1,z)
-G(0,0,1,1,z)
-G(0,1,0,1,z)
-G(0,1,1,0,z)
-G(1,0,0,1,z)
-G(1,0,1,0,z)
-G(1,1,0,0,z)
+3 G(1,1,1,0,z)
-\frac{14 \zeta_{2}^{2}}{5}\,,
\end{autobreak}\\ \nonumber \\
\begin{autobreak}
{\mathcal R}^{(2),{2},\tau(3)}_{\phi\phi\phi} =
4 \zeta_{2} G(1-z,y)
-2 G(0,0,y) G(0,z)
-2 G(0,0,y) G(1,z)
-2 G(0,y) G(0,0,z)
-2 G(0,0,z) G(1-z,y)
+4 G(0,y) G(0,1,z)
+2 G(0,1,z) G(1-z,y)
-4 G(0,1,z) G(-z,y)
+4 G(0,z) G(0,1-z,y)
-2 G(1,z) G(0,z,y)
+4 G(0,y) G(1,0,z)
-2 G(1,0,z) G(1-z,y)
-2 G(0,y) G(1,1,z)
+4 G(1,1,z) G(-z,y)
+4 G(0,z) G(1-z,0,y)
-2 G(1,z) G(1-z,0,y)
-2 G(0,y) G(1-z,1-z,y)
-2 G(0,z) G(1-z,1-z,y)
+4 G(1,z) G(1-z,-z,y)
+4 G(1,z) G(-z,1-z,y)
-4 G(1,z) G(-z,-z,y)
-2 G(0,0,1-z,y)
-2 G(0,1-z,0,y)
-2 G(1-z,0,0,y)
+4 G(1-z,-z,1-z,y)
+4 G(-z,1-z,1-z,y)
-4 G(-z,-z,1-z,y)
+4 \zeta_{2} G(1,y)
+4 G(0,1,0,y)
+4 G(1,0,0,y)
-4 G(1,1,0,y)
+8 \zeta_{2} G(1,z)
-6 G(0,0,1,z)
+2 G(0,1,0,z)
+2 G(0,1,1,z)
+2 G(1,0,0,z)
+2 G(1,0,1,z)
-6 G(1,1,0,z)
+4 \zeta_{3}\,,
\end{autobreak}\\ \nonumber\\
\begin{autobreak}
{\mathcal R}^{(2),{2},\tau(2)}_{\phi\phi\phi} =
-3 G(0,y) G(0,z)
-3 G(0,y) G(1,z)
-3 G(0,y) G(1-z,y)
-2 G(1-z,y)^2
-3 G(0,z) G(1-z,y)
-4 G(1,z) G(1-z,y)
+10 G(1,z) G(-z,y)
+10 G(-z,1-z,y)
-2 G(0,y)^2
+10 G(1,0,y)
-2 G(0,z)^2
-2 G(1,z)^2
+7 G(0,z) G(1,z)
-47 \zeta_{2}
\,,
\end{autobreak}\\ \nonumber\\
\begin{autobreak}
{\mathcal R}^{(2),{2},\tau(1)}_{\phi\phi\phi} =
-10 G(0, y) - 10 G(0, z) - 10 G(1, z) - 10 G(1 - z, y)
\,,
\end{autobreak}\\ \nonumber\\
\begin{autobreak}
{\mathcal R}^{(2),{2},\tau(0)}_{\phi\phi\phi} =
114\,.
\end{autobreak}
\end{align}
%


\subsection{\texorpdfstring{Three-point form factors of energy-momentum tensor operator $\mathcal{O}^3_{\mu\nu}$}{}}
\label{app:results-O3}
As discussed in section~\ref{sec:lt}, the FFs as well as the remainder function of the energy-momentum tensor ${\cal O}^3_{\mu\nu}$ are identical to that of half-BPS:
    \begin{align}
    \label{eq:res-EMTensor}
        &{\cal F}^{(n),{3}}_{ggg,\rm fin} = {\cal F}^{(n),{3}}_{g\phi\phi,\rm fin} = {\cal F}^{(n),{3}}_{g\lambda\bar\lambda,\rm fin}={\cal F}^{(n),{3}}_{\phi\lambda\bar\lambda,\rm fin} \equiv {\cal F}^{(n),{3}, \tau(2n)}_{{\cal C}_1{\cal C}_2{\cal C}_3,\rm fin} = 
        {\cal F}^{(n),\text{half-BPS}}_{{\cal C}_1{\cal C}_2{\cal C}_3,\rm fin}\,,\nonumber\\
        &{\mathcal R}^{(2),{3}}_{ggg} = {\mathcal R}^{(2),{3}}_{g\phi\phi} = {\mathcal R}^{(2),{3}}_{g\lambda\bar\lambda}={\mathcal R}^{(2),{3}}_{\phi\lambda\bar\lambda}\equiv{\mathcal R}^{(2),{3}, \tau(4)}_{{\cal C}_1{\cal C}_2{\cal C}_3} = 
        {\mathcal R}^{(2),\text{half-BPS}}_{{\cal C}_1{\cal C}_2{\cal C}_3} \,.
    \end{align}


\bibliography{n4} 
\bibliographystyle{JHEP}
\end{document}